\documentclass{article}

\usepackage{arxiv}

\usepackage[utf8]{inputenc} 
\usepackage[T1]{fontenc}    
\usepackage{hyperref}       
\usepackage{booktabs}       
\usepackage{nicefrac}       
\usepackage{microtype}      
\usepackage{lipsum}

\usepackage{cite}
\usepackage{amsmath,amssymb,amsfonts}
\usepackage{graphicx}
\usepackage{textcomp}
\usepackage{xcolor}
\usepackage{longtable}
\usepackage{booktabs}
\usepackage{subcaption}
\usepackage{url}
\usepackage{tabularx}
\usepackage{array}
\usepackage{multirow}

\newcommand{\ra}[1]{\renewcommand{\arraystretch}{#1}}
\newcommand{\fourobjects}[3]{%
  \leavevmode\vbox{\vbox{#1}\vspace*{0.02in}\vbox{#2}\vspace*{0.02in}\vbox{#3}}
}

\title{Dehaze-GLCGAN: Unpaired Single Image Dehazing via Adversarial Training}

\author{
  Zahra Anvari \\
  Department of Computer Science and Engineering\\
  University of Texas Arlington\\
  Arlington, TX \\
  \texttt{zahra.anvari@mavs.uta.edu} \\
   \And
 Vassilis Athitsos \\
  Department of Computer Science and Engineering\\
  University of Texas Arlington\\
  Arlington, TX \\
  \texttt{athitsos@uta.edu} \\
}

\begin{document}
\maketitle

\begin{abstract}
Single image dehazing is a challenging problem, and it is far from solved. Most current solutions require paired image datasets that include both hazy images and their corresponding haze-free ground-truth images. However, in reality, lighting conditions and other factors can produce a range of haze-free images that can serve as ground truth for a hazy image, and a single ground truth image cannot capture that range. This limits the scalability and practicality of paired image datasets in real-world applications. In this paper, we focus on \textit{unpaired} single image dehazing and we do not rely on the ground truth image or physical scattering model. We reduce the image dehazing problem to an image-to-image translation problem and propose a dehazing \textbf{G}lobal-\textbf{L}ocal 
\textbf{C}ycle-consistent \textbf{G}enerative \textbf{A}dversarial \textbf{N}etwork (Dehaze-GLCGAN). Generator network of Dehaze-GLCGAN combines an encoder-decoder architecture with residual blocks to better recover the haze free scene. We also employ a global-local discriminator structure to deal with spatially varying haze.
Through ablation study, we demonstrate the effectiveness of different factors in the performance of the proposed network. Our extensive experiments over three benchmark datasets show that our network outperforms previous work in terms of PSNR and SSIM while being trained on smaller amount of data compared to other methods.
\end{abstract}


\section{Introduction}
Haze is an atmospheric phenomenon that can cause visibility issues, and the quality of images captured under haze can be severely degraded. Hazy images suffer from poor visibility and low contrast, which can challenge both human visual perception and numerous intelligent systems relying on computer vision methods. For example, autonomous driving requires the processing of images captured at different weather conditions, and being able to dehaze hazy/foggy images is a must.

The performance of standard computer vision tasks such as object detection~\cite{redmon2016you, liu2016ssd}, semantic segmentation~\cite{long2015fully}, face detection~\cite{yang2016wider, jiang2017face}, clustering and dataset creation~\cite{schroff2015facenet, anvari2019pipeline, lin2018deep,lin2017proximity} can be affected significantly when images are hazy. Hence, image de-hazing can be a useful or even necessary pre-processing task before applying general-purpose computer vision algorithms to hazy images. As a result, single image dehazing has received a great deal of attention over the past ten years.

Most of the recent solutions for image dehazing heavily depend on paired datasets, which include for each hazy image a single clean (haze-free) image as ground truth. In practice, however, there is a range of clean images that can correspond to a hazy image, due to factors such as contrast and light intensity changes throughout the day. In fact, it is infeasible to capture both ground truth/clear image and the hazy image of the same scene at the same time. Thus there is an emerging need to develop solutions that do not rely on the ground truth images and could operate with \textit{unpaired} supervision. 

\begin{figure}
    \centering
    \begin{subfigure}[b]{0.24\textwidth}
        \centering
        \includegraphics[width=0.85\textwidth]{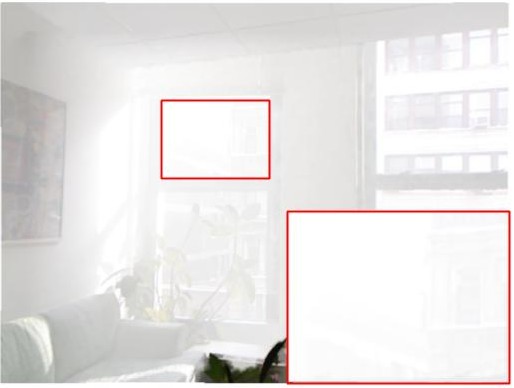}
        \caption[hazy]%
        {{\small Hazy}}    
        \label{fig:hazy}
    \end{subfigure}
    \hspace*{-0.1in}%
    \begin{subfigure}[b]{0.24\textwidth}  
        \centering 
        \includegraphics[width=0.85\textwidth]{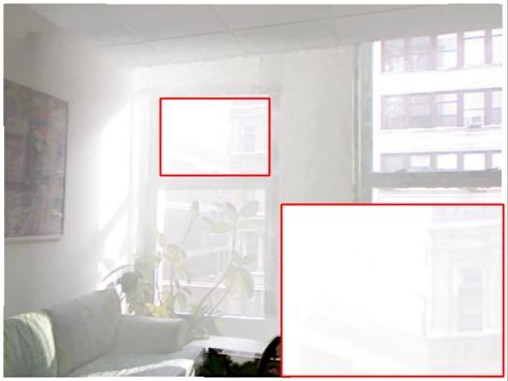}
        \caption[]%
        {{\small AOD-Net (ICCV'17)}}    
        \label{fig:aodnet}
    \end{subfigure}
    \begin{subfigure}[b]{0.24\textwidth}   
        \centering 
        \includegraphics[width=0.85\textwidth]{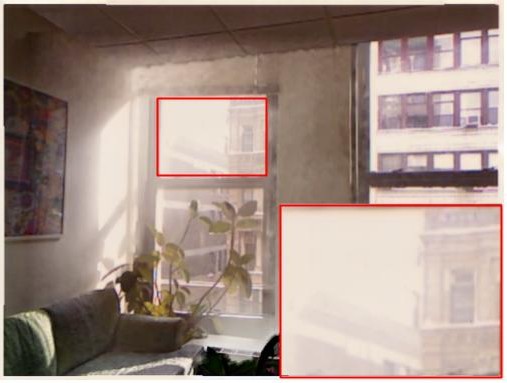}
        \caption[]%
        {{\small EPDN (CVPR'19)}}    
        \label{fig:epdn}
    \end{subfigure}
    \hspace*{-0.1in}%
    \begin{subfigure}[b]{0.24\textwidth}   
        \centering 
        \includegraphics[width=0.85\textwidth]{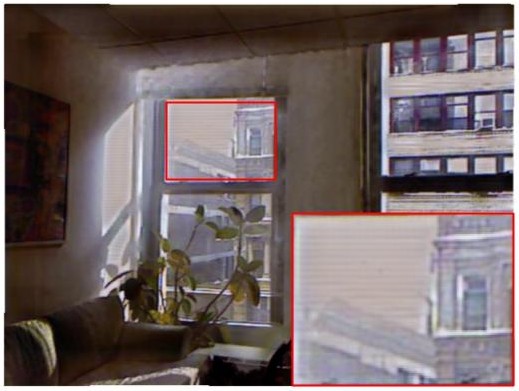}
        \caption[]%
        {{\small Ours}}    
        \label{fig:ours1}
    \end{subfigure}
    \caption{A single image dehazing example. Our method generates an image with less haze and rich details compared with AOD-Net~\cite{li2017aod} and EPDN~\cite{qu2019enhanced}.}
    {\small } 
    \label{fig:comparison}
\end{figure}

~\textit{Unpaired image dehazing} does not rely on the ground truth of hazy images, meaning that there is no one-to-one mapping between hazy and haze-free images in the training dataset. Thus it is more scalable and more practical. In this paper, we focus on unpaired image dehazing and propose a model to remove haze from a hazy image with spatially varying haze conditions.




Single image dehazing methods can be categorized into two main classes: prior-based~\cite{ancuti2016night,ancuti2010fast,emberton2015hierarchical,meng2013efficient,tarel2009fast} and learning-based~\cite{cai2016dehazenet,zhang2017joint,yang2018towards,swami2018candy,ren2016single}. Prior-based models solve the haze removal problem through estimating the physical model,~\emph{i.e}. transmission map and atmospheric light parameters. Learning-based methods mainly use CNN-based or GAN-based models to recover the haze-free/clean image. These models take advantage of large amounts of training data to learn a model that maps a hazy image to a clean image.

In this paper, we treat the unpaired image dehazing as an image-to-image translation problem. We propose a novel cycle-consistent generative adversarial network with novel generators, that operates without paired supervision and benefits from (i) a global-local discriminator architecture to handle spatially varying haze (ii) customized cyclic perceptual loss to generate more realistic and natural images. Through empirical analysis we show that the proposed network can effectively remove haze while being trained on smaller amounts of data compared to other methods.

Figure \ref{fig:comparison} shows the result of our network compared to the current state-of-the-art methods. As we can see, the proposed method (Figure \ref{fig:ours1}) removes haze more effectively and generates a more realistic clean image compared to previous work, while being trained on a smaller training dataset compared to previous work.

In summary, this paper presents the following contributions:
\begin{itemize}
    \item We propose a novel cycle-consistent generative adversarial network called Dehaze-GLCGAN for unpaired image dehazing. Dehaze-GLCGAN does not rely on the physical scattering model, as opposed to many previous methods, and instead it adopts the image-to-image translation approach for unpaired image dehazing ~\cite{engin2018cycle,zhu2017unpaired}.\\
    \item We propose a novel dehazing generator that combines an encoder decoder structure with embedded Residual blocks to better preserve the details of images.
    \item We employ color loss along with perceptual loss to generate more visually pleasing images.
    \item We adopt a global-local discriminator structure to deal with spatially varying haze and generate cleaner images.
    \item Through empirical analysis, we show that our network outperforms the previous work in terms of PSNR and SSIM with a large margin while trained on smaller amount of data compared to previous work.
\end{itemize}

This paper is organized as follows. Section 2 describes background information as well as previous work. Section 3 describes our proposed method. Section 4 presents our experiments and results.

\section{Related Work}
Single image dehazing and generative adversarial networks are the two topics that are the most relevant for our method. We briefly describe them and their related work in this section.

\subsection{Single image dehazing}
Numerous attempts have been done to solve the single image haze removal problem. These methods can be categorized into two main classes: prior-based and learning-based, that we describe them below.

\subsubsection{Prior-based dehazing}
Prior-based methods are mainly based on prior information and assumptions to recover the haze-free images from hazy images. They heavily depend on estimating the the parameters of the physical scattering model~\cite{mccartney1976optics,srinivasa2002vision},~\emph{aka.} the atmospheric scattering model, which contains the transmission map and the atmospheric light to solve the haze removal problem. The physical scattering model is formulated as:
\begin{equation}\label{eq_phy1}
I(x) = J(x)t(x) + A(1 - t(x))
\end{equation}
where $I(x)$ is the hazy image, $J(x)$ is the haze-free image or the scene radiance, $t(x)$ is the medium transmission map, and $A$ is the global atmospheric light on each $x$ pixel coordinates. 
He~\emph{et al.}~\cite{he2010single} proposed a dark channel prior to estimate the transmission map effectively. Tan et al.~\cite{tan2008visibility} increase the contrast of hazy images, based on the fact that haze-free images have higher contrast than hazy images.

\subsubsection{Learning-based dehazing}
Recently learning based methods have been proposed that utilize CNNs and GANs for the single image dehazing problem. CNN-based methods try to recover the clean images through the atmospheric scattering model, by mainly estimating the transmission map and atmospheric light~\cite{mccartney1976optics,narasimhan2000chromatic}.

MSCNN~\cite{ren2016single} contains two sub-networks called coarse-scale and fine-scale, to estimate the transmission map. The coarse-scale network estimates the transmission map and is further improved locally by the fine-scale network.
In DehazeNet~\cite{cai2016dehazenet}, authors modified the classic CNN model by adding feature extraction and non-linear regression layers. These modifications distinguish DehazeNet from other CNN-based models. The All-In-One Dehazing Network (AOD-net)~\cite{li2017aod} proposed an end-to-end network that produces the haze-free/clean images through reformulating the atmospheric scattering model.
DCPDN~\cite{zhang2018densely} uses an end-to-end single image dehazing network to jointly learn the transmission map, atmospheric light and image dehazing. EPDN~\cite{qu2019enhanced} employs a three-part network which consists of generator, discriminator, and two enhancer blocks. 



\subsection{Generative Adversarial Networks}
GANs have become one of the most successful methods for image generation, manipulation, restoration, and reconstruction. GANs have been used to super-resolve images~\cite{ledig2017photo}, remove motion blurriness from images~\cite{kupyn2018deblurgan}, and remove noise~\cite{chen2018image}, to name a few applications. GANs are also utilized in image dehazing~\cite{yang2018towards,engin2018cycle}. Yang et al.~\cite{yang2018towards} proposed the disentangled dehazing network without paired supervision.The GAN network they proposed consists of three generators: a generator for generating haze-free image, a generator for the atmospheric light, and the third generator for transmission map.

The Cycle-consistency GAN (CycleGAN)~\cite{zhu2017unpaired} method was proposed for unpaired image-to-image translation task and has gained significant attention during the past couple of years. Engin et al.~\cite{engin2018cycle} utilized CycleGAN for image dehazing along with the perceptual loss to generate more visually realistic dehazed images.  


\section{Proposed Method}
We first reduce the unpaired image dehazing problem to an image-to-image translation problem, and then propose a dehazing \textbf{G}lobal-\textbf{L}ocal \textbf{C}ycle-consistent \textbf{G}enerative \textbf{A}dversarial \textbf{N}etwork (Dehaze-GLCGAN) with discriminators that operate at global and local levels, to translate a hazy image to a haze-free image. Next we describe our network in detail.


\begin{figure*}
     \centering
     \begin{subfigure}[b]{1.0\textwidth}
        \centering
        \includegraphics[width=0.90\textwidth]{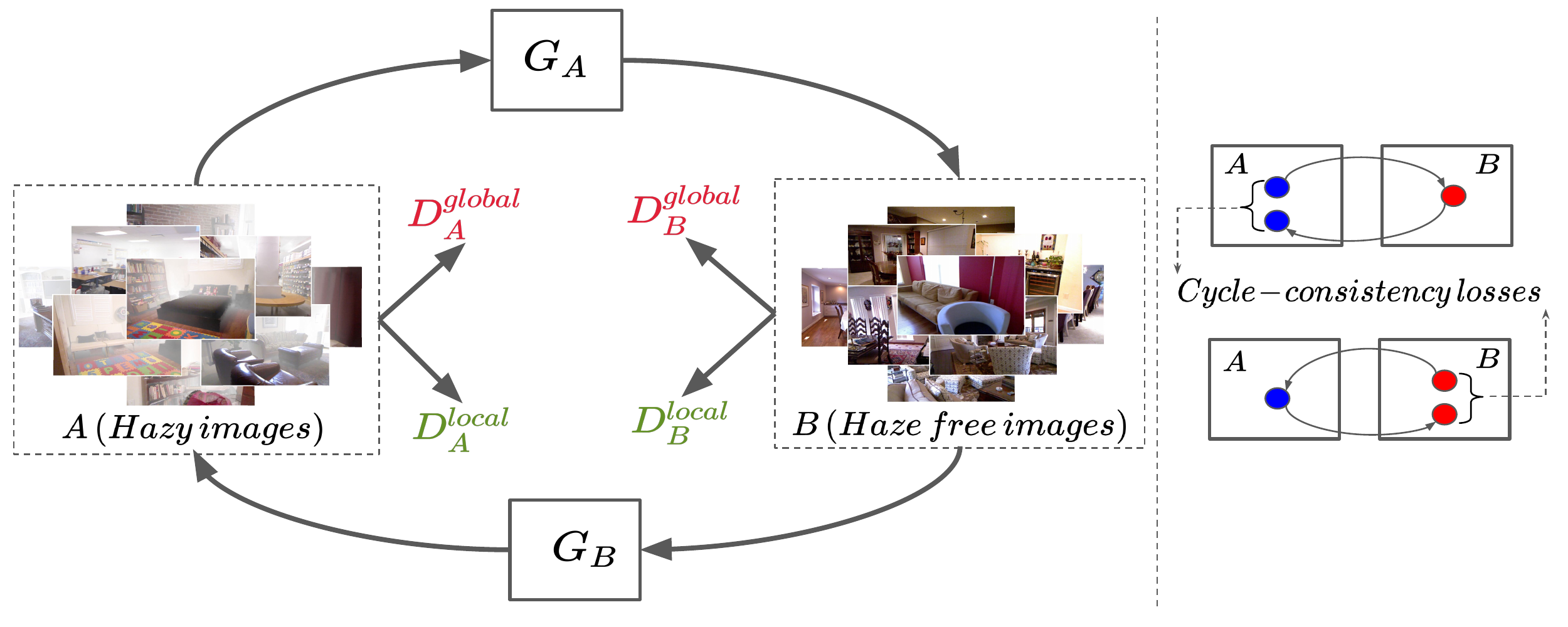}
        \caption{An overview of Dehaze-GLCGAN}
        \label{fig:mcdgan}
     \end{subfigure}
     \begin{subfigure}[b]{1.0\textwidth}
        \centering
        \includegraphics[width=6in,height=3.5in]{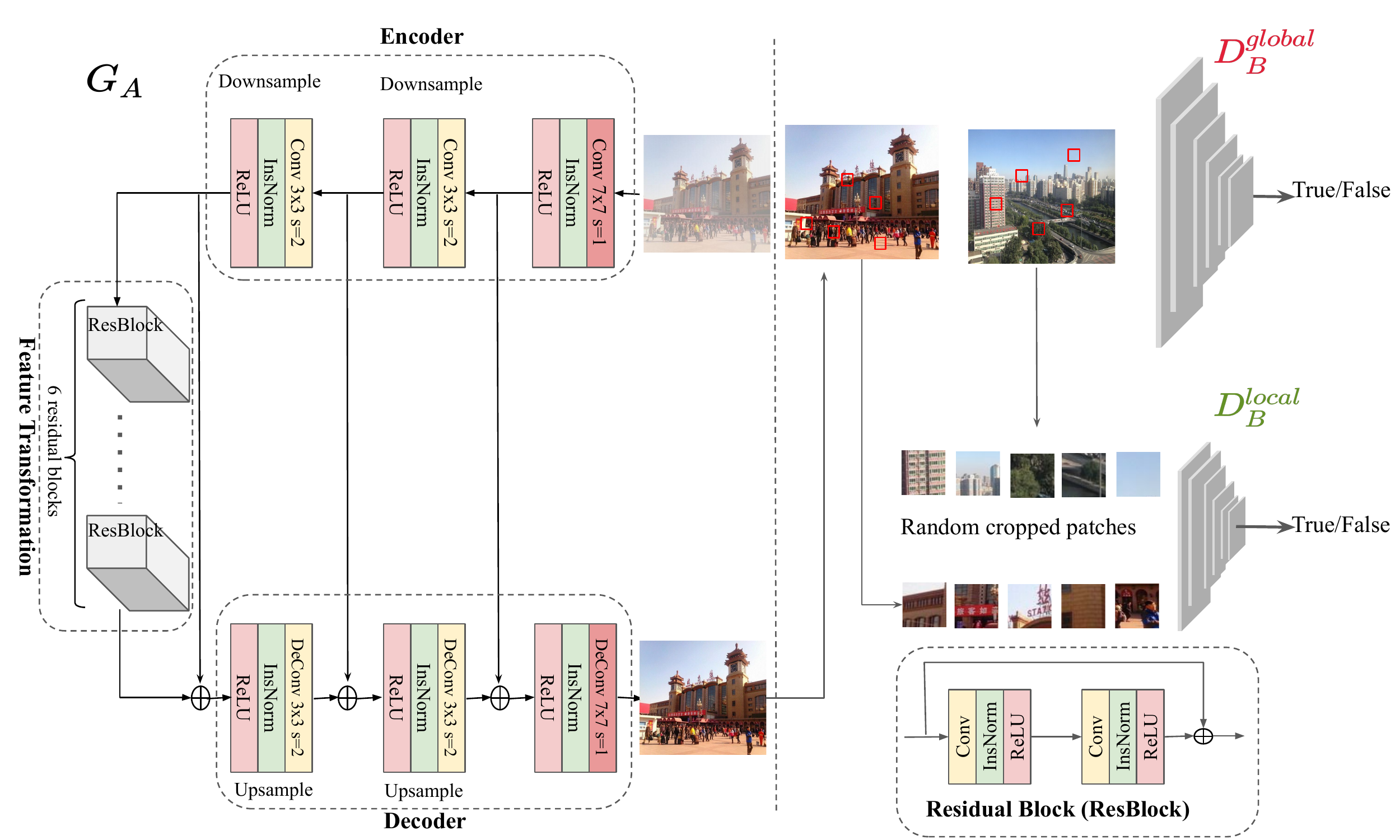}
        \caption{The architecture of Dehaze-GLCGAN. This figure shows the architecture of $G_A$, $D_B^{Global}$ and $D_B^{Local}$. $G_B$, $D_A^{Global}$ and $D_A^{Local}$ have the same architecture as $G_A$, $D_B^{Global}$, $D_B^{Local}$ respectively, except that they work on different inputs, ~\emph{e.g.,} the input to $G_B$ is a clean image and the input to $G_A$ is a hazy image.}
        \label{fig:generatorDescriminator}
     \end{subfigure}
        \caption{The overview and architecture of Dehaze-GLCGAN}
        \label{fig:three graphs}
\end{figure*}

\subsection{Overview of Dehaze-GLCGAN}
Figure~\ref{fig:mcdgan} demonstrates an overview of the proposed network, and Figure~\ref{fig:generatorDescriminator} depicts the architecture of the generators and discriminators. The bottom right part of Figure~\ref{fig:generatorDescriminator} shows in more detail the structure of the residual blocks that are used in the network.

This network is designed to translate images from domain $A$,~\emph{i.e.} hazy, to domain $B$,~\emph{i.e.} clean/haze-free, in a cycle-consistent manner. The components of Dehaze-GLCGAN are Generator $A$ ($G_A$), Generator $B$ ($G_B$), Global Discriminator $A$ ($D_A^{Global}$), Global Discriminator $B$ ($D_B^{Global}$), Local Discriminator $A$ ($D_A^{Local}$), and Local Discriminator $B$ ($D_B^{Local}$).

Figure \ref{fig:mcdgan} demonstrates how the generators and discriminators interact. $G_A$ translates images from domain $A$ (hazy) to domain $B$ (haze-free). Global discriminator $B$ ($D_B^{Global}$) classifies if a haze-free image generated by $G_A$ is real or fake, \textbf{based on the entire image}. Local discriminator $B$ ($D_B^{Local}$) classifies if a haze-free image generated by $G_A$ is real or fake, \textbf{based on 5 randomly cropped image patches of size $64 \times 64$ pixels from that image} (inspired by~\cite{jiang2019enlightengan}).

Similarly, $G_B$ translates images from domain $B$ (haze-free) to domain $A$ (hazy). And $D_A^{Global}$ and $D_A^{Local}$ classify if a hazy image generated by $G_B$ is real or fake based on the entire image and 5 randomly cropped patches respectively.






$G_A$ and $G_B$ utilize the same network architecture (Figure~\ref{fig:three graphs}). Also all discriminators share the same network architecture, however operate on different scales. 
Through ablation study we show that utilizing global-local discriminators effectively improves the network performance. 

\subsection{Generator}
Figure~\ref{fig:generatorDescriminator} presents the architecture of Dehaze-GLCGAN. The architecture of generator $A$ ($G_A$) is depicted on the left. Note that $G_B$ has the same architecture. In order to generate a haze-free image without paired supervision in a cycle-consistent manner, we require a generator network that can preserve the images' texture, structure and details while removing haze. To do so, we designed a network with three modules: encoder, feature transformation, and decoder. 

The encoder module starts with a convolution layer followed by an Instance Normalization and Relu non-linearity and then two downsampling blocks. Our next module, feature transformation, uses 6 Residual Blocks to extract more complex and deeper features whilst removing haze. The main benefit of going deeper in network is that it becomes capable of representing very complex functions and also learns features at many different levels of abstraction. 

The third module, which is the decoder, consists of two upsampling blocks which are deconvolution layers, followed by Instance Normalization and Relu. The deconvolution layers of the decoder are used to recover image structural details and convert the feature maps to a haze-free RGB image. Through the deconvolution layer also known as transposed layer, the upsampling operations are performed to obtain intermediate feature mappings with double spatial size and half channels than its previous counterpart.

\subsection{Discriminator}
The right side of Figure~\ref{fig:generatorDescriminator} shows $D_B^{Global}$ and $D_B^{Local}$. Note that $D_A^{Global}$ and $D_A^{Local}$ have the same architecture as $D_B^{Global}$ and $D_B^{Local}$ respectively. We have two types of discriminators, global and local, each performing a particular operation to classify real vs. fake images. Initially our model contained only global discriminators. However, we have observed that global discriminators often fail on spatially-varying haze images,~\emph{i.e.}, in cases where haze density variation exists in an image, and thus different image parts need to be enhanced differently from other parts.
In order to enhance each local region appropriately, in addition to improving the haze removal globally, we utilized a global-local discriminator scheme inspired by~\cite{jiang2019enlightengan} in a cycle-consistent manner, both using PatchGAN discriminators.

PatchGAN discriminators classify \textbf{individual (N x N) patches} in the image as real vs. fake, as opposed to classifying the \textbf{entire image} as real vs. fake. 

PatchGAN imposes more constraints which help with sharp high-frequency details. Furthermore, PatchGAN has fewer number of parameters thus performs faster compared to classifying the entire image~\cite{isola2017image}. It is shown that $70 \times 70$ patches produce the best results~\cite{isola2017image}, thus we employed the same patch size in our network.


\subsection{Loss functions}
Our objective loss function contains: i) adversarial losses for matching the distribution of generated images to the data distribution in the target domain; ii) cycle consistency losses to prevent the learned mappings $G_A$ and $G_B$, as mentioned in Fig. \ref{fig:mcdgan}, from contradicting each other, and iii) cyclic perceptual loss to help the generators generate more visually pleasing images. Next we describe these loss functions in detail.

\subsubsection{Adversarial loss}
Here we adopted Least Squares GAN (LSGAN)~\cite{mao2017least} to calculate the adversarial loss. It has been shown that LSGAN is able to generate higher quality images than regular GANs (\emph{aka.} Vanilla GANs)~\cite{mao2017least}. In addition, it is more stable during the learning process. Equations~\ref{eq3} and~\ref{eq4} show how we calculate the adversarial loss for the global discriminator and the global generator respectively.

\begin{equation}\label{eq3}
\begin{split}
L_D^{Global} = E_{x_r \sim P_{real}} [(D(x_r) - 1)^2] +
E_{x_f \sim P_{fake}}  [(D(x_f) - 0])^2]
\end{split}
\end{equation}
\begin{equation}\label{eq4}
L_G^{Global} = E_{x_r \sim P_{fake}} [(D(x_f) - 1)^2]
\end{equation}

where $D$ denotes the discriminator, and $x_r$ and $x_f$ are sampled from the real and fake distribution respectively.

We introduced the local discriminator to enhance small regions of the hazy image. Equations~\ref{eq5} and~\ref{eq6} depicts the corresponding loss functions:
\begin{equation}\label{eq5}
\begin{split}
L_D^{Local} = E_{x_r \sim P_{real-patches}} [(D(x_r) - 1)^2] + 
E_{x_f \sim P_{fake-patches}} [(D(x_f) - 0)^2]
\end{split}
\end{equation}
\begin{equation}\label{eq6}
L_G^{Local} = E_{x_f \sim P_{fake-patches}} [(D(x_f) - 1)^2]
\end{equation}

where $D$ denotes the discriminator, $x_r$ and $x_f$ are sampled from the ~\textbf{patches} taken from real and fake distribution.

\subsubsection{Cycle consistency loss}
Adversarial loss can not guarantee that the learned function can map an individual input $x_i$ to desired output $y_i$. Thus a cycle-consistency loss is proposed by CycleGAN to reduce the space of possible mapping functions. Cycle-consistency loss function ($L1-norm$) compares the cyclic image and the original image in unpaired image-to-image translation process~\cite{zhu2017unpaired}. Cycle consistency loss can be written as:

\begin{equation}
\begin{split}
L_{cycle}(G_A,G_B) = E_{x \sim p_{data(x)}}[\left\Vert(G_B(G_A(x)) - x)\right\Vert]_1 +  E_{y \sim p_{data(y)}}[\left\Vert(G_A(G_B(y)) - y)\right\Vert]_1
\end{split}
\end{equation}
where $G_A$ and $G_B$ are forward and backward generators, $x$ belongs to domain $X$ (\emph{i.e.} the original domain, hazy images here) and $y$ belongs to domain $Y$ (\emph{i.e.} the haze-free images). $G_B(G_A(x))$ and $G_A(G_B(y))$ are the reconstructed images. 

If cycle-consistency loss's goal is met, the reconstructed images $G_B(G_A(x))$ will match closely to the input image $x$ and also the reconstructed images $G_A(G_B(y))$ will match closely to the input image $y$.

\subsubsection{Color loss}
Hazy images usually lack brightness and contrast, to improve these lacking features we employed color loss~\cite{ignatov2017dslr} to measure the color difference between the enhanced images and haze-free images. This loss function forces the generator to generate images with the same color distribution as the haze-free images.

To implement this loss function we apply a Gaussian blur to the enhanced image and the haze-free image and then we compute the Euclidean distance between the blurred images. 
\begin{equation}
L_{color}(A, B) = \left\Vert(A_{blurred} - B_{blurred})\right\Vert_2^2
\end{equation}

$A_{blurred}$ and $B_{blurred}$ are the blurred images of A and B images.

\subsubsection{Cyclic perceptual loss}
Adversarial and cycle consistency loss are not able to preserve the textual and perceptual information of corrupted hazy images. Therefore, to achieve the perceptual similarity we employed a cyclic perceptual loss inspired by cycle consistency loss, to make sure that this information is recovered as much as possible. 

This loss can be calculated using a pre-trained VGG16~\cite{simonyan2014very} model to calculate the feature space distance between images. This loss function has been employed by other vision-based tasks~\cite{ledig2017photo,kupyn2018deblurgan}. It constrains the extracted feature distance between the output image and its ground truth. 

Since our method is designed using unpaired supervision,~\emph{i.e.} the ground truth of images is not available, perceptual loss cannot be directly applied. Thus a modified version of perceptual loss is adopted when the ground truth is unavailable, originally proposed by~\cite{jiang2019enlightengan}. They calculate the loss function between the original image and its enhanced version and call it self preserving perceptual loss and show that it helps with preserving the image structure and details. We adopted this loss and utilized it in a cycle-consistent manner, and we call it Cyclic Perceptual Loss.


The goal of this loss function is to preserve the image structure and content features during dehazing. 
To calculate this loss, we focused on feature maps extracted from the $2^{nd}$ and $5^{th}$ pooling layers of VGG-16 pre-trained model. Equation~\ref{eq7} shows how this loss is calculated:
\begin{equation}\label{eq7}
Loss_{CP}(I_h) = \frac{1}{W_{i,j}H_{i,j}} \sum_{x=1}^{W_{i,j}} \sum_{y=1}^{H_{i,j}} (\sigma_{i,j}(I_h) - \sigma_{i,j}(G(I_h)))^2 
\end{equation}
where $I_h$ represents the hazy input image and $G(I_h)$ represents the dehazed output. Symbol $\sigma_{i,j}$ denotes the features extracted from the pre-trained (on ImageNet dataset) VGG16 model. $W_{i,j}$ and $H_{i,j}$ are the dimensions of the extracted feature. $i$ shows the $i_{th}$ max pooling and $j$ denotes the $j_{th}$ conv layer after $i_{th}$ max pooling layer. 

We also add an instance normalization layer~\cite{ulyanov2017improved} after the VGG feature maps before feeding into $L_{CP}$ and $L^{Local}_{CP}$ in order to stabilize training. To calculate the $L^{Local}_{CP}$ for the local discriminator we used the cropped local patches of input and output images and used the same equation~\ref{eq7}.

\subsection{Final objective function}
The overall loss function for training Dehaze-GLCGAN is defined as follows:
\begin{equation}\label{eq8}
\begin{split}
Loss_{total} = L_{global}^{GAN} + L_{local}^{GAN} + L_{global}^{Cycle} + L_{local}^{Cycle} +  L_{global}^{CP} + L_{local}^{CP} + L_{global}^{color} + L_{local}^{color}
\end{split}
\end{equation}

\section{Experiments and Results}

Previous \textit{paired} methods have primarily reported results on the ITS dataset~\cite{li2018benchmarking}. To compare with those methods (even though our method is unpaired), we evaluated our method on the ITS dataset. However, we should stress that, due to lack of computing power, \textbf{we only train our model over 20\% of the ITS training set, while previous methods are trained over the entire ITS dataset.} The ITS dataset contains 100,000 indoor hazy images. This dataset is part of the RESIDE dataset~\cite{li2018benchmarking}, which is a standard dataset of synthetic hazy images. RESIDE also has a test dataset called SOTS which contains 500 indoor hazy images and 500 outdoor hazy images with their clear ground truth. We report results on the SOTS test dataset.

To evaluate the performance of our method compared to previous \textit{unpaired} methods, we trained a model over NYU-Depth dataset~\cite{ancuti2016d}, and test it on the NYU-Depth and Middlebury datasets. We should note that \textbf{previous unpaired methods were trained over the augmented NYU-Depth dataset, while we do not use the augmented data}. The NYU-Depth and Middlebury datasets belong to the D-hazy dataset~\cite{ancuti2016d}, which is a standard synthetic hazy dataset for image dehazing. NYU-Depth contains 1449 hazy images couples with their ground truth images and Middlebury has 23 high-resolution (2k images) hazy images with their ground truth. NYU-Depth has a ground truth image for each hazy image. Since our method uses unpaired supervision, the training process received no information about which haze-free image corresponds to which hazy image.  

\subsection{Training}
For training we need two sets of training data: trainA includes hazy images and trainB includes ground truth images (shuffled to simulate the unpaired supervision like other unpaired methods~\cite{yang2018towards}). We opted for Adam optimizer ($momentum = 0.5$) with batch size of 1. Our initial learning rate ($\alpha$) was set to $0.0002$ for 100 epochs, with linear decay to zero over the next 100 epochs. We implemented our model in PyTorch using eight NVIDIA Tesla P100 GPUs and trained our network for 200 epochs.

\subsection{Quality Measures}
We used the following measures, to analyze the performance of our proposed method:
\begin{itemize}
    \item \textbf{PSNR (Peak Signal to Noise Ratio)}: measures the ratio between the maximum possible value of a signal and the power of distorting noise that affects the quality of its representation. The higher the PSNR, the better the reconstructive method. 
    
    \item \textbf{SSIM (Structural Similarity Index)}: is a perceptual metric that quantifies image quality degradation caused by processing. 
    In this measurement, image degradation is considered as the change of perception in structural information~\cite{kumar2013visual}.
    
    
    \item \textbf{PI (Perceptual Index)}: measures the quality of restored and reconstructed images based on human perception~\cite{blau20182018}. A lower perceptual index indicates better perceptual quality. 
    \item \textbf{CIEDE2000}: This measures the color difference between hazy and dehazed images; smaller values indicate better color preservation, thus better dehazing and perceptual quality~\cite{luo2001development}.
\end{itemize}

\subsection{Ablation Study}
To demonstrate the effectiveness of the local discriminator and cyclic perceptual loss, we conduct several ablation experiments. In particular, to evaluate the effectiveness of local discriminators, we trained our model without the local discriminators and tested the model on NYU-Depth dataset. Afterwards, to demonstrate the effect of the perceptual loss, we trained a model without this loss function and evaluated the model on NYU-Depth dataset. 

\begin{figure}
    \centering
    \begin{subfigure}[b]{0.24\textwidth}
        \centering
        {\footnotesize \textbf{\textcolor{red}{PSNR: 12.14}, \textcolor{violet}{SSIM: 0.74}}}
        \includegraphics[width=0.99\textwidth]{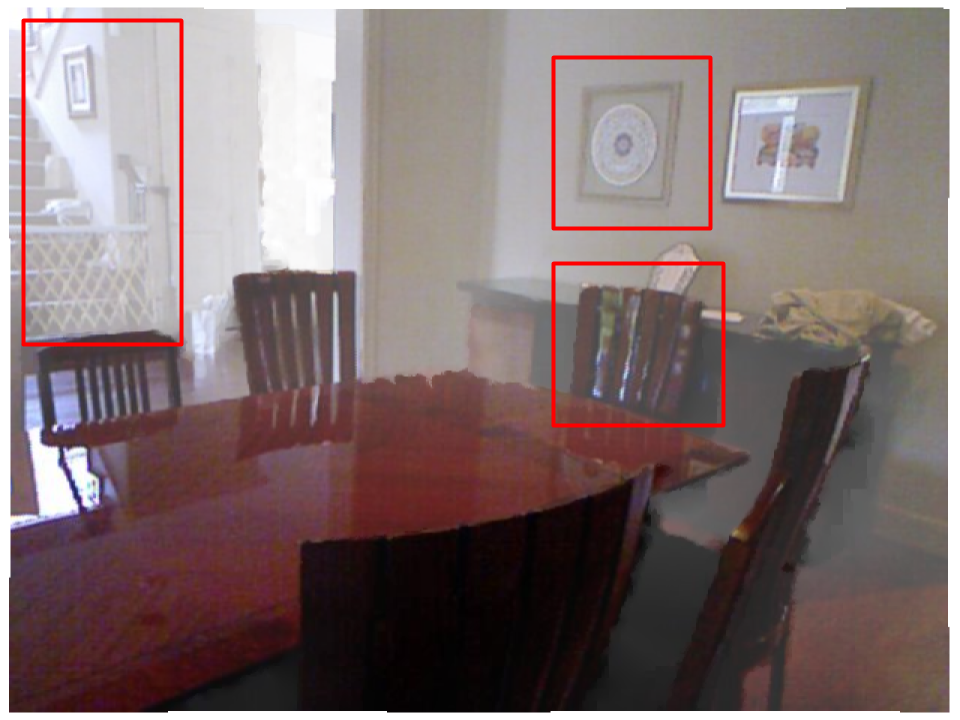}
        \caption[hazy]%
        {Hazy image}    
        \label{fig:hazy}
    \end{subfigure}
    \hspace*{-0.1in}%
    \begin{subfigure}[b]{0.24\textwidth}  
        \centering 
        {\footnotesize \textbf{\textcolor{red}{PSNR: 13.43}, \textcolor{violet}{SSIM: 0.78}}}
        \includegraphics[width=0.99\textwidth]{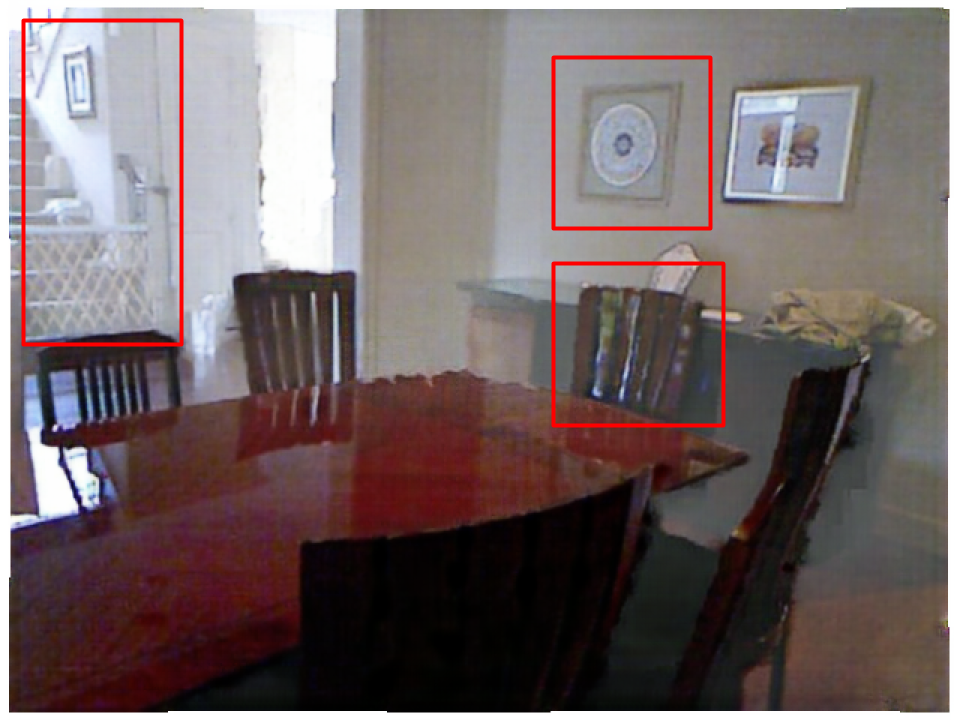}
        \caption[]%
        {CycleGAN}    
        \label{fig:aodnet}
    \end{subfigure}
    \begin{subfigure}[b]{0.24\textwidth}   
        \centering 
        \includegraphics[width=0.99\textwidth]{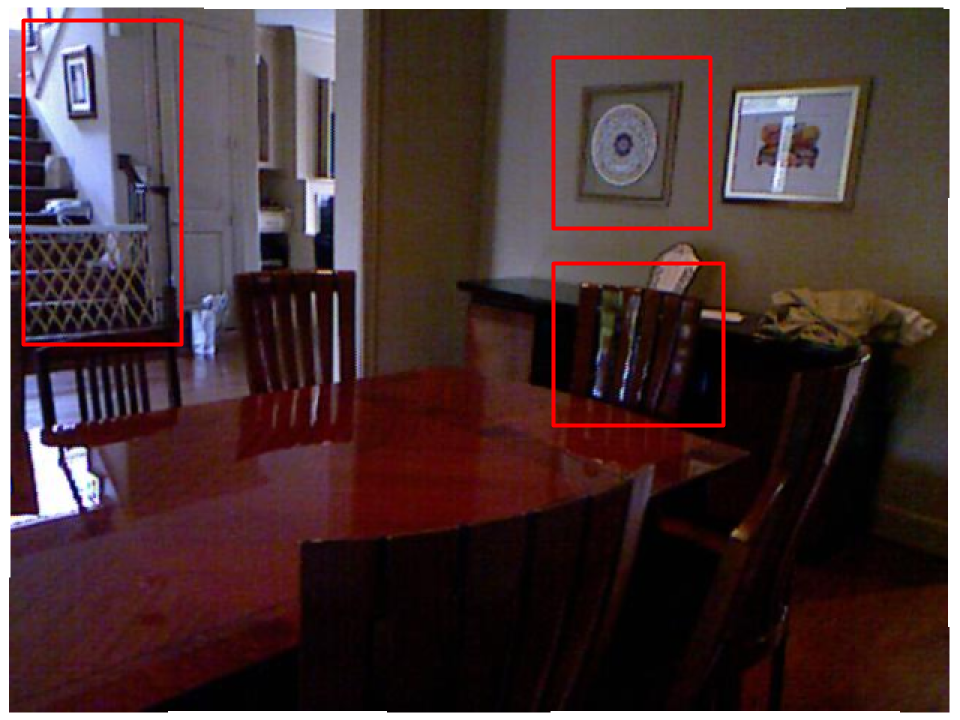}
        \caption[]%
        {Haze free}    
        \label{fig:epdn}
    \end{subfigure}
    \hspace*{-0.1in}%
    \begin{subfigure}[b]{0.24\textwidth}   
        \centering 
        {\footnotesize \textbf{\textcolor{red}{PSNR: 19.11}, \textcolor{violet}{SSIM: 0.90}}}
        \includegraphics[width=0.99\textwidth]{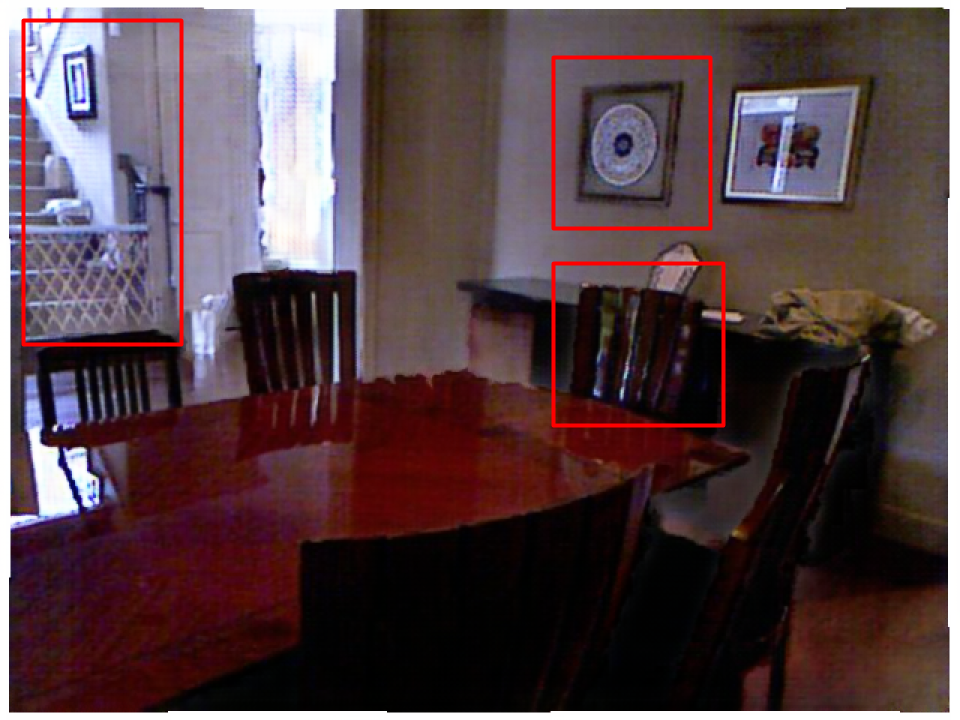}
        \caption[]%
        {Ours}    
        \label{fig:ours1}
    \end{subfigure}
    \caption{Comparison between CycleGAN and the proposed method.}
    {\small } 
    \label{fig:comparison_cyclegan}
\end{figure}

\begin{table*}
\ra{1.2}
\centering
\caption{Ablation study over NYU-Depth dataset. The larger values of PSNR, SSIM and the smaller value of CIEDE2000 indicate the better dehazing and perceptual quality.}
\begin{tabular}{llll}\hline
 \textbf{Setting} & \textbf{$\boldsymbol{\uparrow}$ PSNR} & \textbf{$\boldsymbol{\uparrow}$SSIM} & \textbf{$\boldsymbol{\downarrow}$ CIEDE2000} \\\hline
  CycleGAN & 13.3879 & 0.5223 & 17.6113 \\
  Dehaze-GLCGAN w/o color loss & 14.5402 & 0.7407 & 15.6401 \\ 
  Dehaze-GLCGAN w/o perceptual loss & 14.6582 & 0.7312 & 15.6348 \\ 
  Dehaze-GLCGAN w/o residual blocks & 14.1092 & 0.6923 & 16.4344 \\ 
  Dehaze-GLCGAN w/o local discriminator & 14.0681 & 0.7111 & 19.9466 \\
  Dehaze-GLCGAN & \textbf{15.4780} & \textbf{0.7808} & \textbf{14.8877} \\ 
\end{tabular}
\label{table:ablation}
\end{table*}

Table \ref{table:ablation} depicts the results of our ablation study in terms of PSNR, SSIM, and CIEDE2000. It is observed that incorporating local discriminators can help achieve better PSNR, SSIM and CIEDE2000, meaning better restoration and generation of more visually pleasing results.

Having removed cyclic perceptual and color loss, our model achieved lower PSNR, SSIM, and CIEDE2000, compared to its counterpart,~\emph{i.e.} perceptual loss and color incorporated. This means our generator can generate more natural and perceptually pleasing clean images when the cyclic perceptual and color loss function is utilized. The best results in terms of PSNR, SSIM, and CIEDE2000 are achieved when both local discriminators and cyclic perceptual loss are incorporated.

Figure~\ref{fig:comparison_cyclegan} also shows the comparison of the proposed method with CycleGAN. As we can see, the proposed method can achieve higher PSNR and SSIM compared to CycleGAN and removes more hazy.

\subsection{Quantitative and Qualitative Analysis}
\noindent\textbf{Comparison with paired methods:} To evaluate the performance of our proposed method quantitatively, we first compare our model with {\em paired} methods on the SOTS test dataset, since that is the dataset where existing paired methods have reported results. As noted earlier, for these experiments, our method as well as the competitors are trained on the ITS dataset, however our method was only trained on 20\% of that dataset.

\begin{table}
\ra{1.2}
\centering
\caption{Results on SOTS outdoor dataset. Most of the numbers for the previous work are taken from~\cite{qu2019enhanced}. }
\begin{tabular}{llll}\hline
 Method &$\boldsymbol{\uparrow}$PSNR & $\boldsymbol{\uparrow}$SSIM & $\boldsymbol{\downarrow}$PI\\\hline
DCP~\cite{he2010single} & 19.13 & 0.8148 & 2.5061\\
DehazeNet~\cite{cai2016dehazenet} & 22.46 & 0.8514 & 2.4346\\
AOD-NET~\cite{li2017aod} & 20.29 & 0.8765 &{2.4280}\\
DCPDN~\cite{zhang2018densely} & 19.93 & 0.8449 & 2.7269\\
EPDN~\cite{qu2019enhanced} & 22.57 & 0.8630 & \textbf{2.3858}\\
\textbf{Ours} & \textbf{23.0276} & \textbf{0.9165} &  2.4051\\
\end{tabular}
\label{table:sots_results}
\end{table}

\begin{table}
\ra{1.2}
\centering
\caption{Results on NYU-Depth dataset. The numbers for the previous work are taken from~\cite{yang2018towards,engin2018cycle}.}
\begin{tabular}{llll}\hline
 Method & $\boldsymbol{\uparrow}$PSNR & $\boldsymbol{\uparrow}$SSIM & $\boldsymbol{\downarrow}$CIEDE2000\\\hline
CycleGAN~\cite{cai2016dehazenet} & 13.3879 & 0.5223 & 17.6113\\
Cycle-Dehaze~\cite{li2017aod} & 15.41 & 0.66 & NA\\
DDN~\cite{he2010single} & 15.5456 & 0.7726 &  11.8414\\
\textbf{Ours} & \textbf{15.9780} & \textbf{0.8208} & \textbf{10.3862}\\
\end{tabular}
\label{table:nyu_results}
\end{table}

\begin{table}
\ra{1.2}
\centering
\caption{Results on Middlebury dataset. The numbers for the previous work are taken from~\cite{yang2018towards,engin2018cycle}.}
\begin{tabular}{llll}\hline
 Method & $\boldsymbol{\uparrow}$PSNR & $\boldsymbol{\uparrow}$SSIM & $\boldsymbol{\downarrow}$CIEDE2000\\\hline
CycleGAN~\cite{cai2016dehazenet} & 11.3037 & 0.3367 & 26.3118\\
Cycle-Dehaze~\cite{li2017aod} & 12.7634 & 0.5812 & NA\\
DDN~\cite{he2010single} & 14.9539 & 0.7741 &  \textbf{13.4826}\\
\textbf{Ours} & \textbf{15.6802} & \textbf{0.8482} &  16.6745\\
\end{tabular}
\label{table:middleberry_results}
\end{table}

\begin{figure*}
  \begin{subfigure}{0.166\textwidth}
  \centering
          \fourobjects
  {\includegraphics[width=0.99\textwidth]{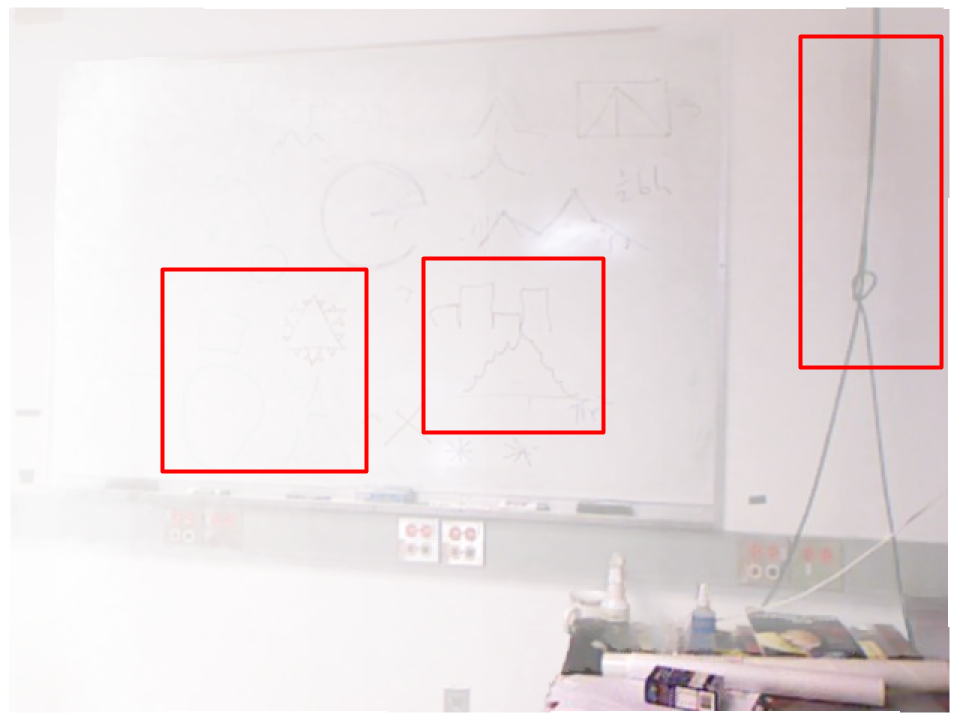}}
  {\includegraphics[width=0.99\textwidth]{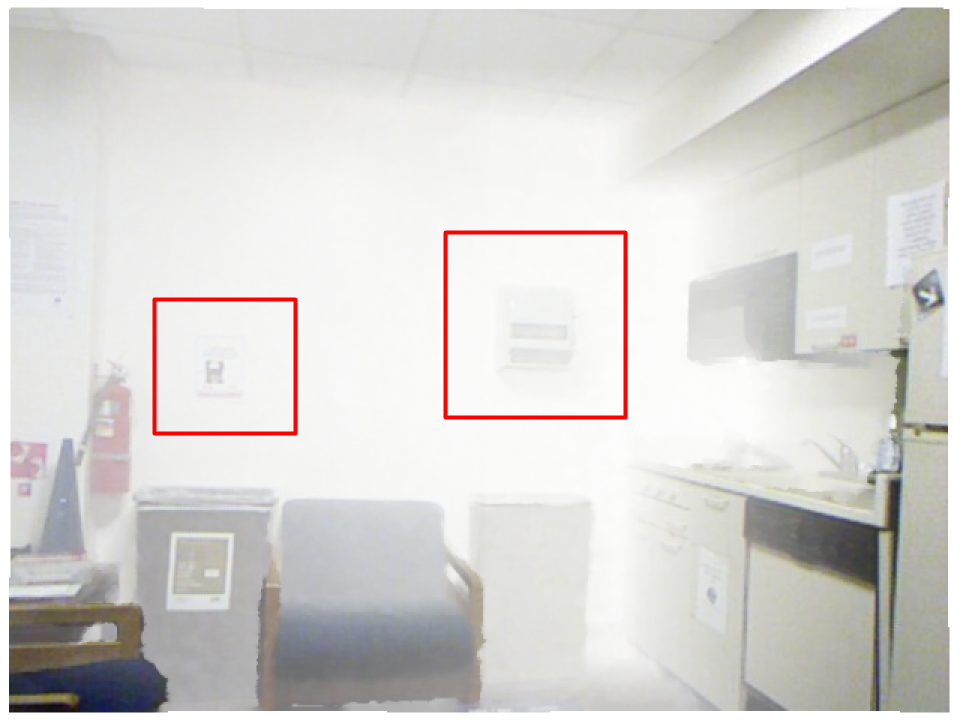}}
  {\includegraphics[width=0.99\textwidth]{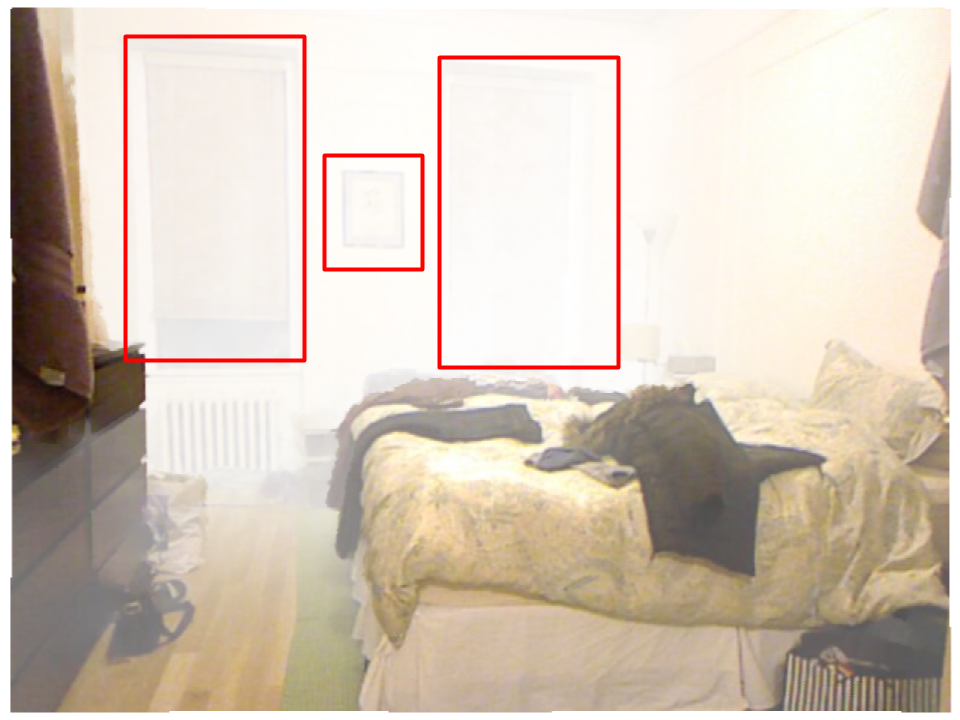}}
    \caption{Hazy}
    \label{fig:hazy}
  \end{subfigure}
\hspace*{-0.04in}%
  \begin{subfigure}{0.166\textwidth}
  \centering
          \fourobjects
  {\includegraphics[width=0.99\textwidth]{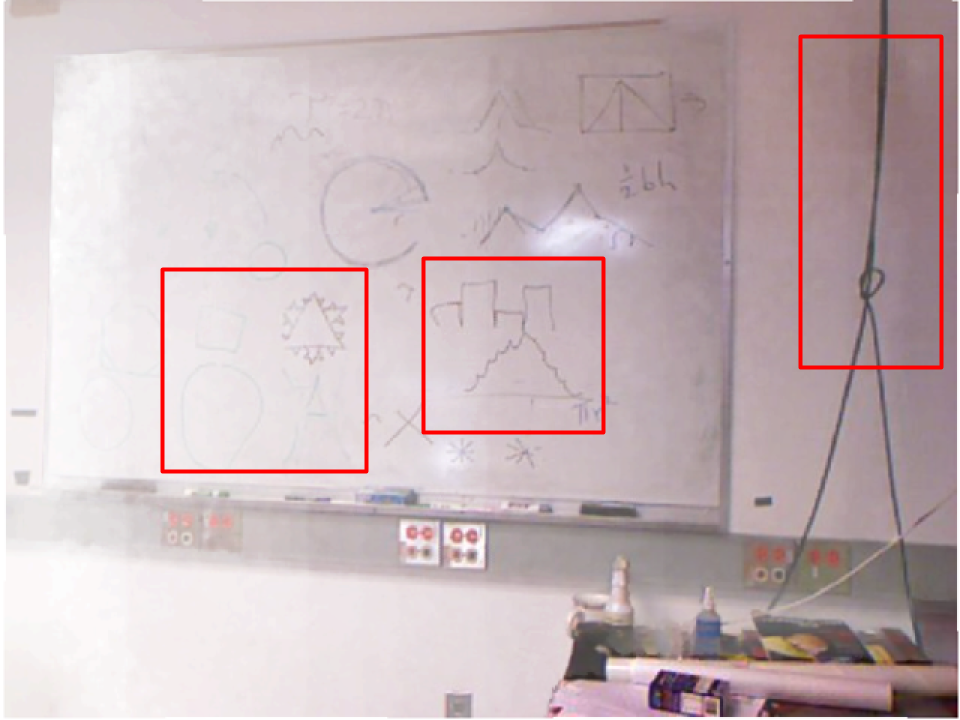}}
  {\includegraphics[width=0.99\textwidth]{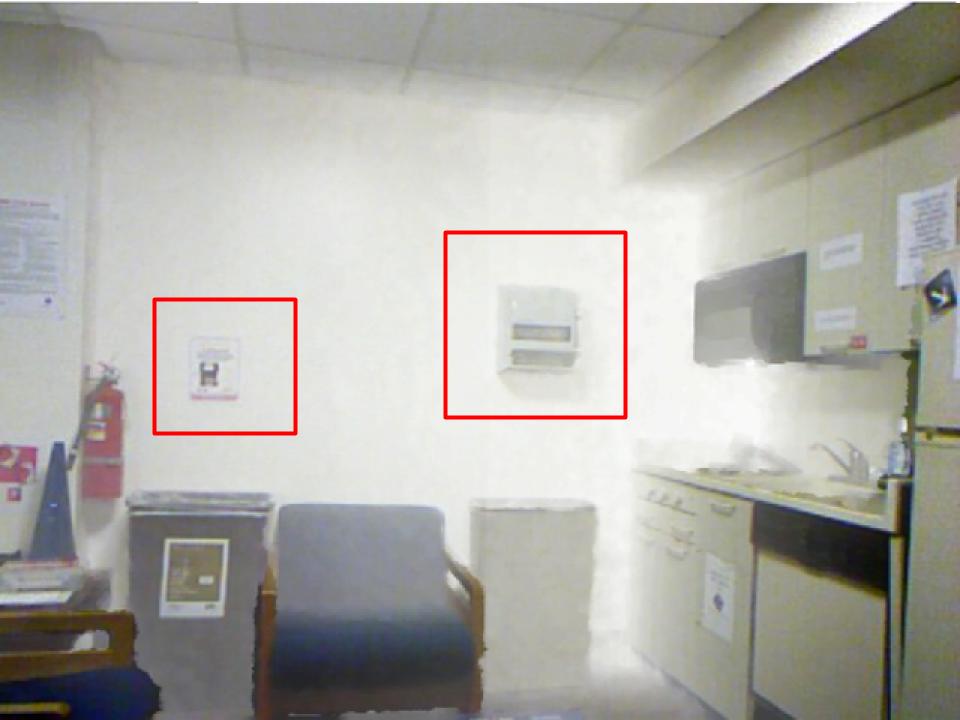}}
  {\includegraphics[width=0.99\textwidth]{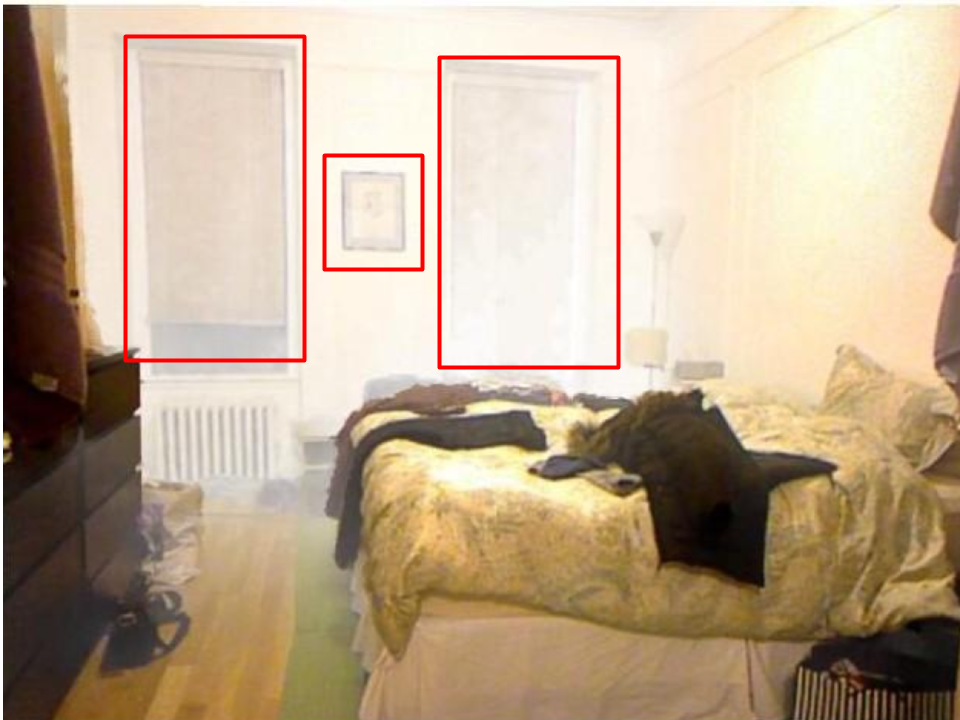}}
    \caption{DehazeNet}
    \label{fig:mscnn}
  \end{subfigure}
\hspace*{-0.04in}%
  \begin{subfigure}{0.166\textwidth}
  \centering
          \fourobjects
  {\includegraphics[width=0.99\textwidth]{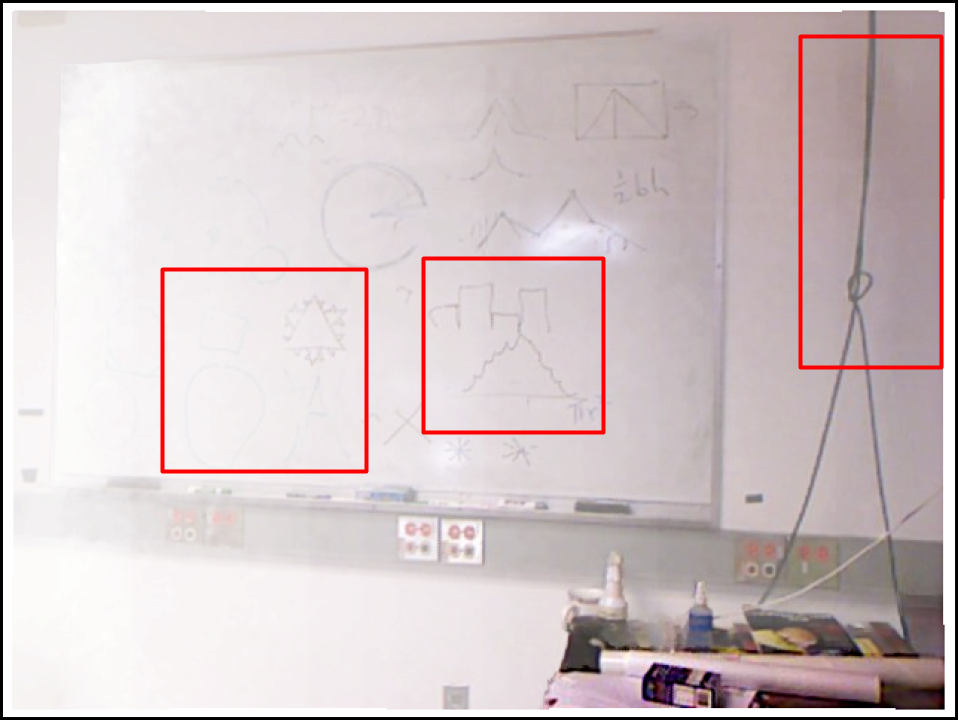}}
  {\includegraphics[width=0.99\textwidth]{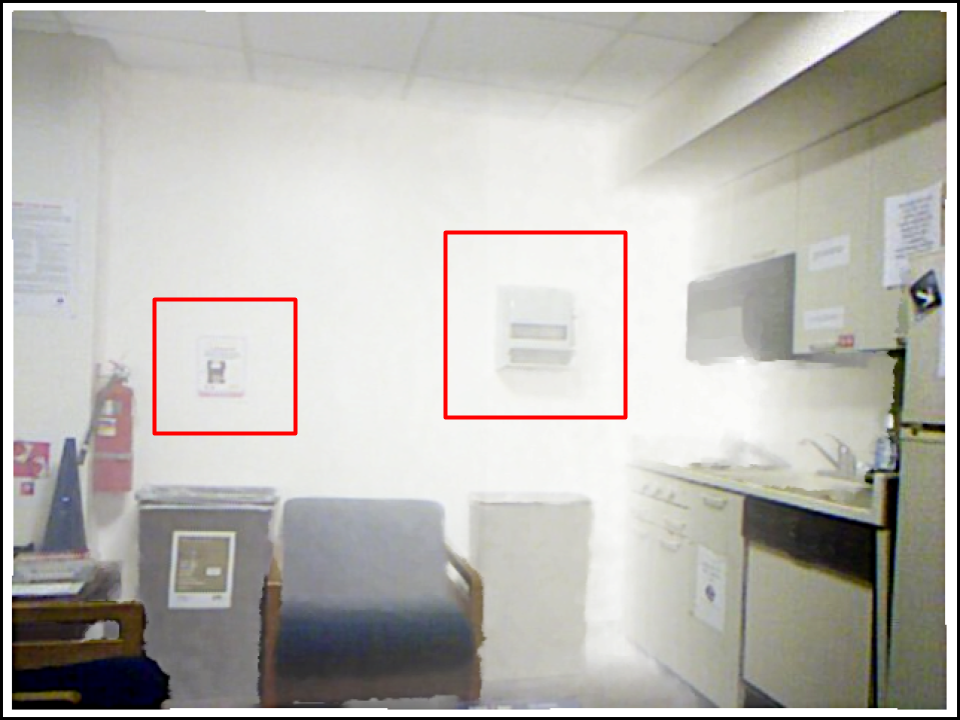}}
  {\includegraphics[width=0.99\textwidth]{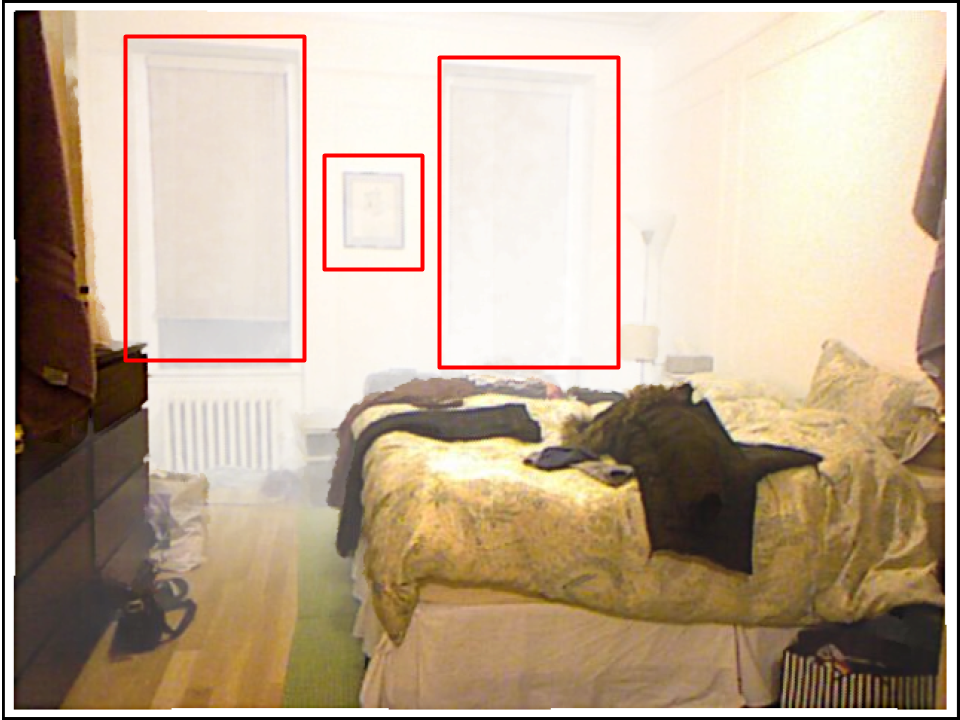}}
    \caption{AOD-Net}
    \label{fig:aodnet}
  \end{subfigure}
\hspace*{-0.04in}%
  \begin{subfigure}{0.166\textwidth}
  \centering
          \fourobjects
  {\includegraphics[width=0.99\textwidth]{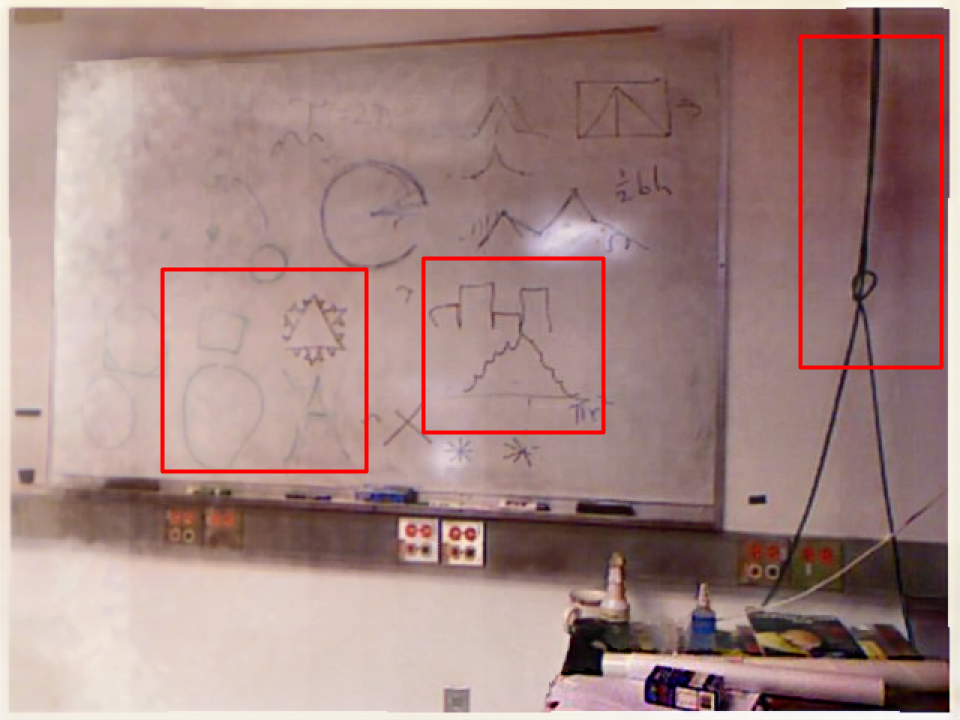}}
  {\includegraphics[width=0.99\textwidth]{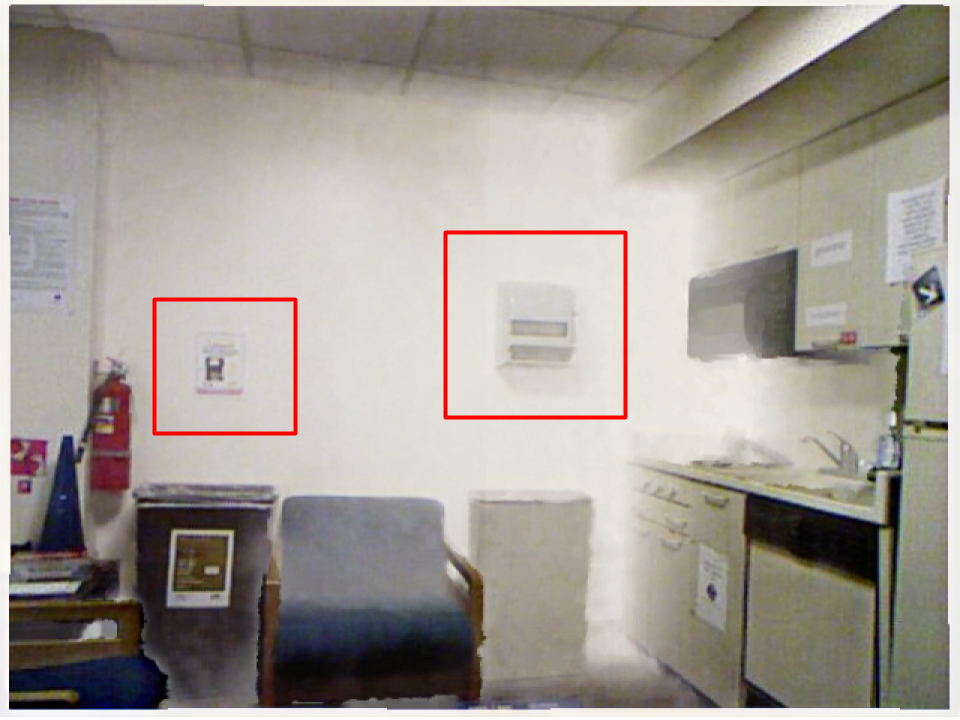}}
  {\includegraphics[width=0.99\textwidth]{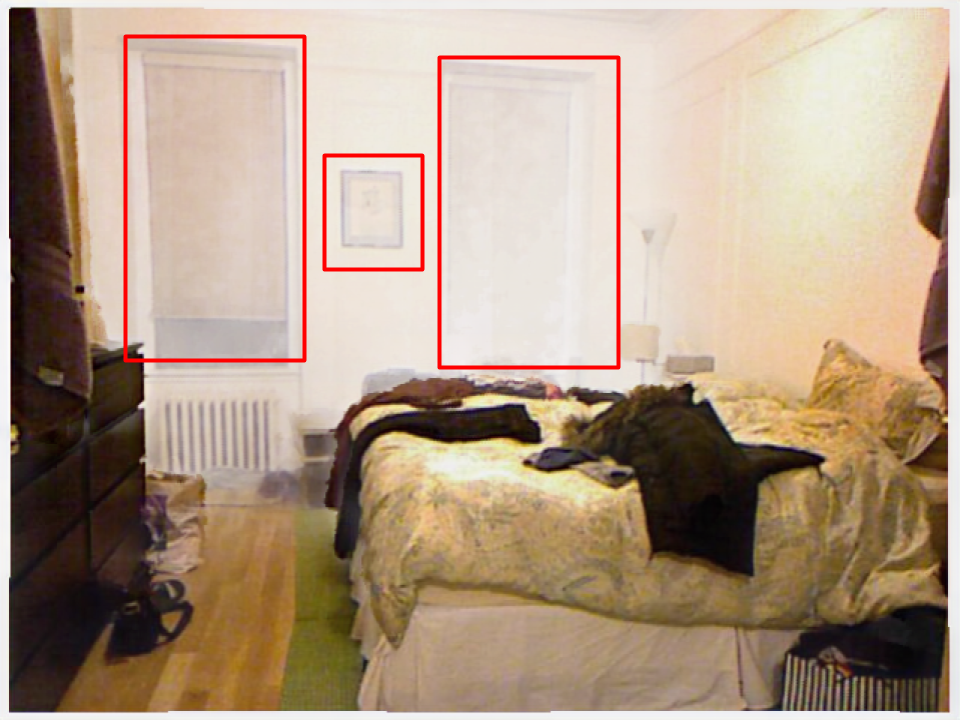}}
    \caption{EPDN}
    \label{fig:dcpdn}
  \end{subfigure}
 \hspace*{-0.04in}%
  \begin{subfigure}{0.166\textwidth}
  \centering
          \fourobjects
  {\includegraphics[width=0.99\textwidth]{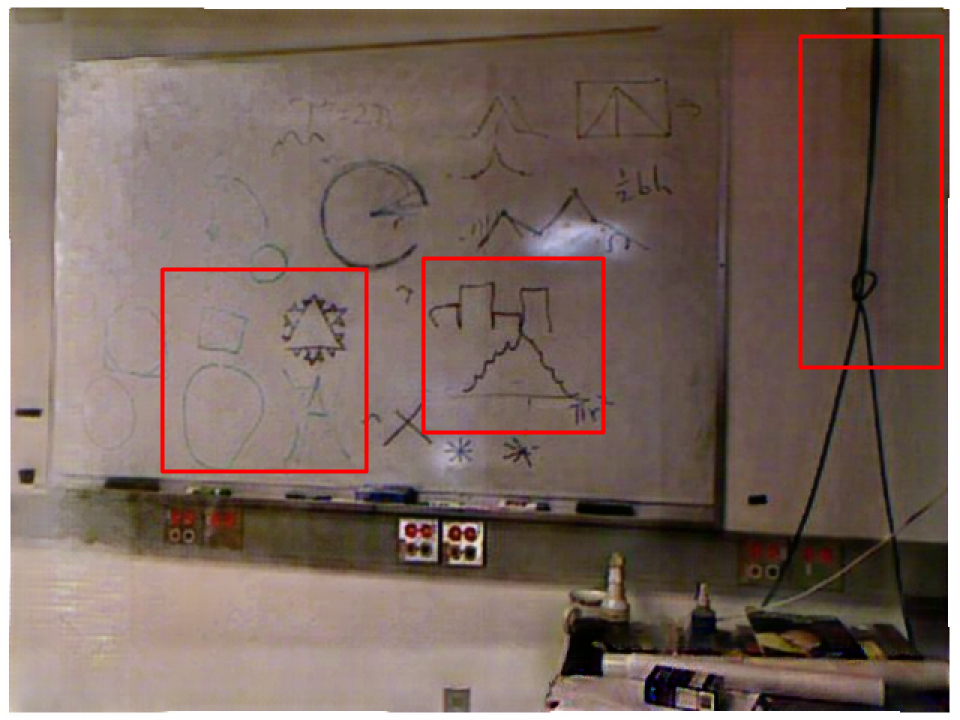}}
  {\includegraphics[width=0.99\textwidth]{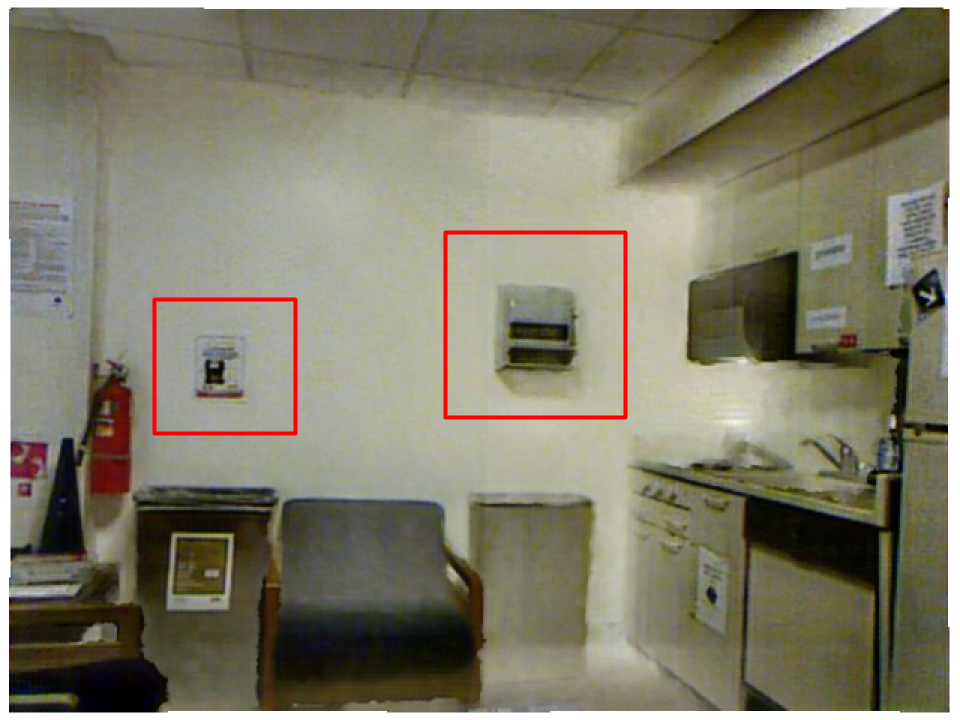}}
  {\includegraphics[width=0.99\textwidth]{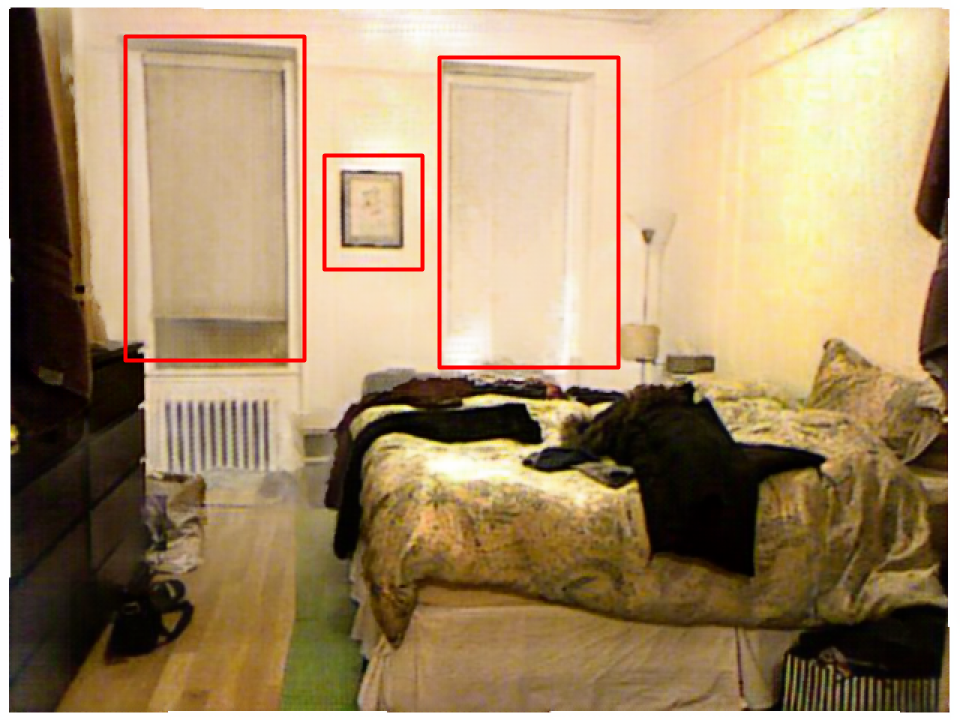}}
    \caption{Ours}
    \label{fig:ours}
  \end{subfigure}
  \hspace*{-0.04in}%
  \begin{subfigure}{0.166\textwidth}
  \centering
          \fourobjects
  {\includegraphics[width=0.99\textwidth]{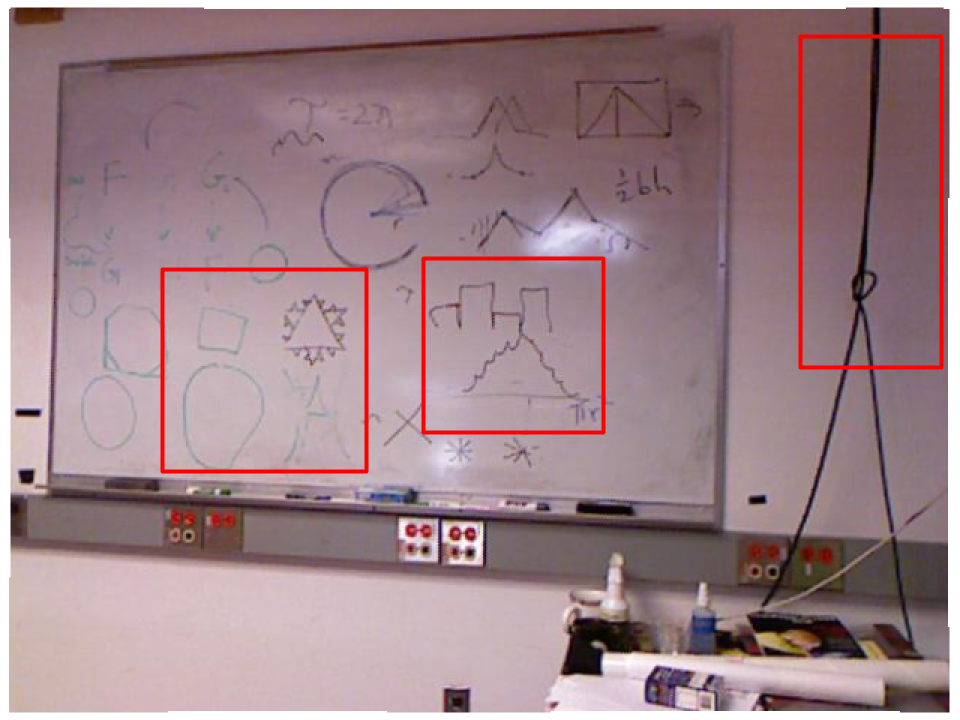}}
  {\includegraphics[width=0.99\textwidth]{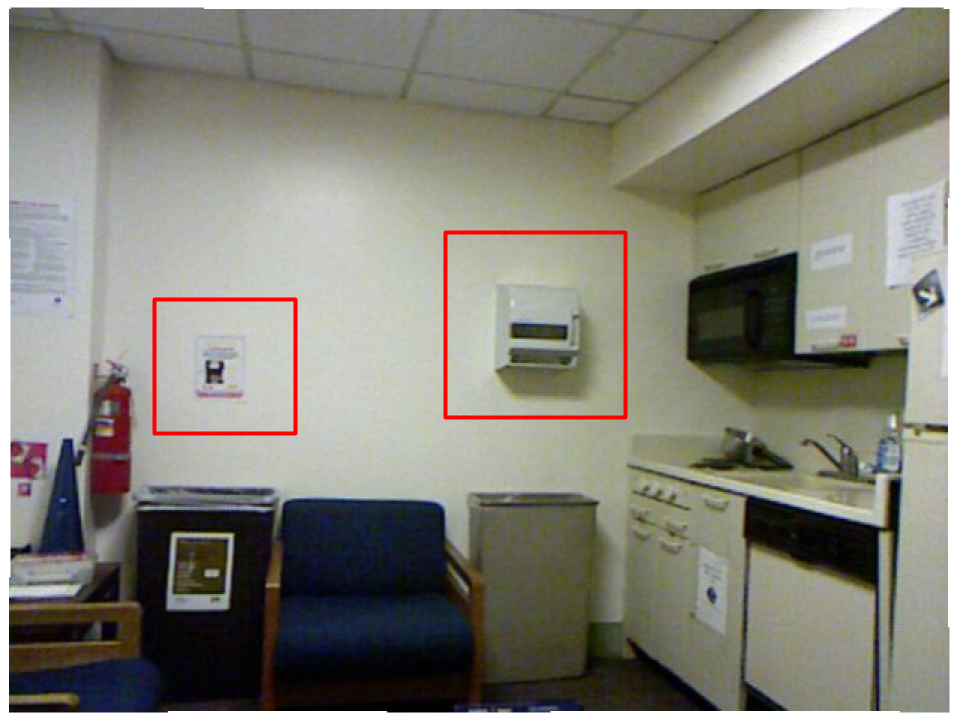}}
  {\includegraphics[width=0.99\textwidth]{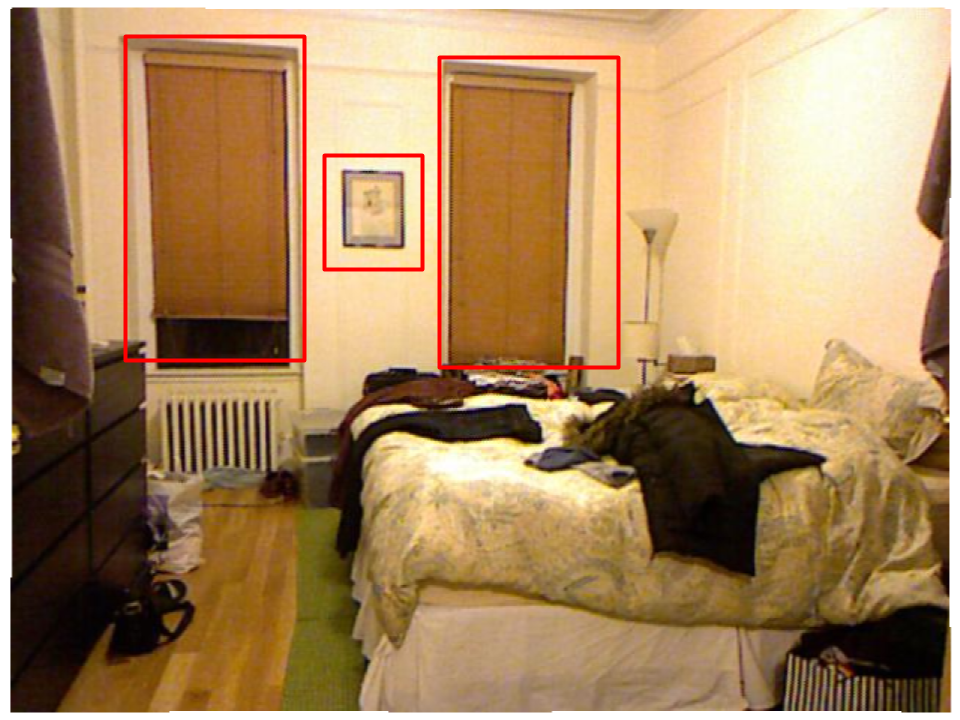}}
    \caption{Haze free}
    \label{fig:gt}
  \end{subfigure}
  \caption{Comparison of the state-of-the-art dehazing methods on NYU-Depth dataset. Please magnify to see the details. For high resolution version of these images, please refer to the supplementary material.}
  \label{fig:examples_nyu}
\end{figure*}

Table~\ref{table:sots_results} shows the results on SOTS test dataset. It demonstrates that Dehaze-GLCGAN achieves the best performance in single image dehazing in terms of both PSNR and SSIM. These results were achieved despite the fact that our method is unpaired, and the competitors use paired supervision.

\noindent\textbf{Comparison with unpaired methods:} We also compare our model with {\em unpaired} methods, on the NYU-Depth and Middlebury datasets. As noted earlier, for these experiments, our method as well as the competitors are trained on the NYU-Depth dataset, however the competitors used an augmented version of that dataset, whereas we used the original, unaugmented version. Table~\ref{table:nyu_results} and ~\ref{table:middleberry_results} show the results on NYU-Depth and Middle-berry datasets respectively. Our method outperforms the other methods in terms of PASNR and SSIM. 


\begin{figure*}
  \begin{subfigure}{0.142\textwidth}
  \centering
          \fourobjects
  {\includegraphics[width=0.95\textwidth]{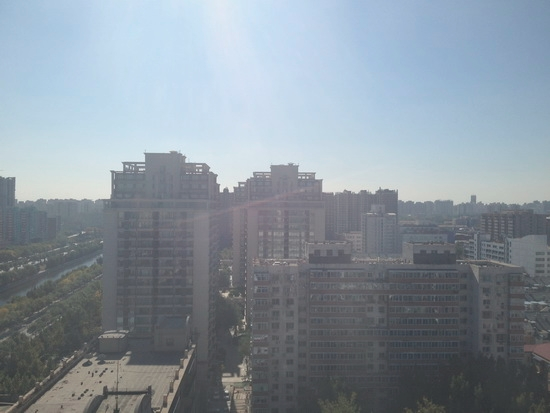}}
  {\includegraphics[width=0.95\textwidth]{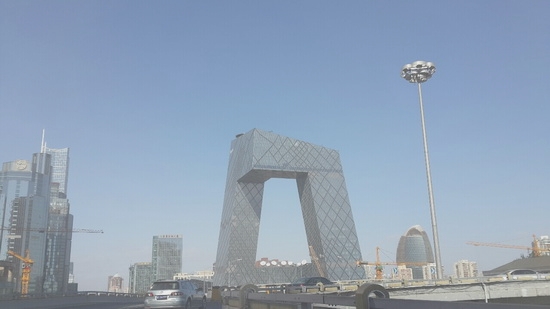}}
  {\includegraphics[width=0.95\textwidth]{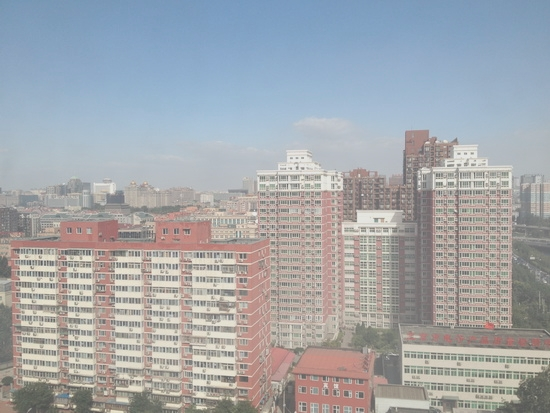}}
    \caption{Hazy}
    \label{fig:hazy}
  \end{subfigure}
  \hspace*{-0.04in}%
  \begin{subfigure}{0.142\textwidth}
  \centering
          \fourobjects
  {\includegraphics[width=0.95\textwidth]{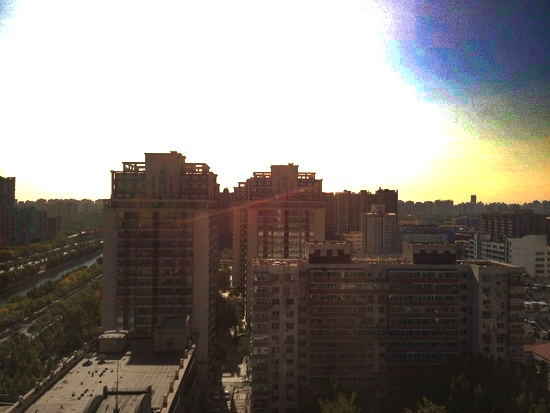}}
  {\includegraphics[width=0.95\textwidth]{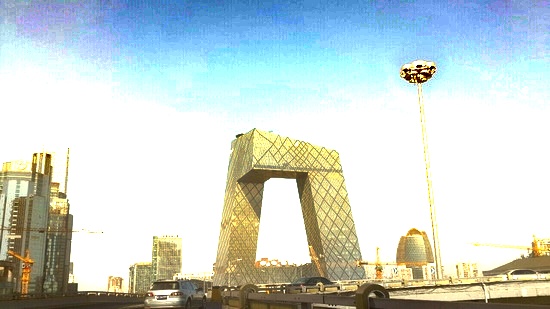}}
  {\includegraphics[width=0.95\textwidth]{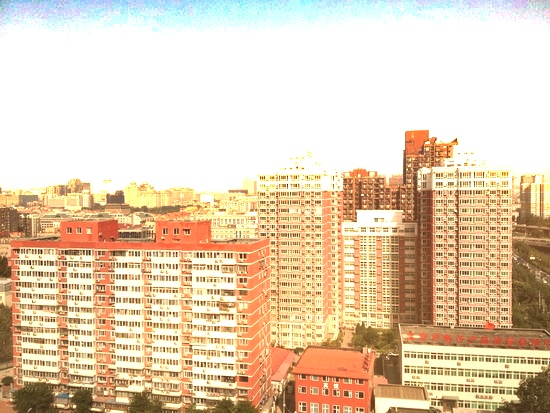}}
    \caption{DCP}
    \label{fig:dcp}
  \end{subfigure}
\hspace*{-0.04in}%
  \begin{subfigure}{0.142\textwidth}
  \centering
          \fourobjects
  {\includegraphics[width=0.95\textwidth]{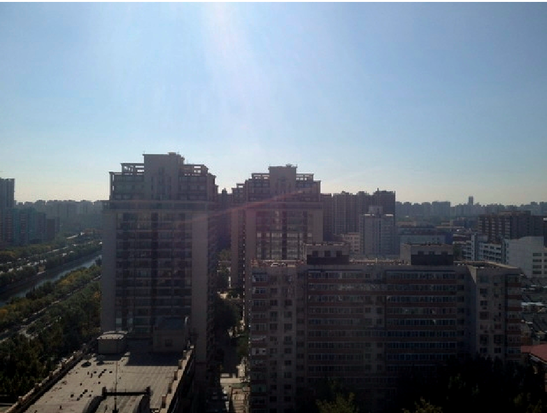}}
  {\includegraphics[width=0.95\textwidth]{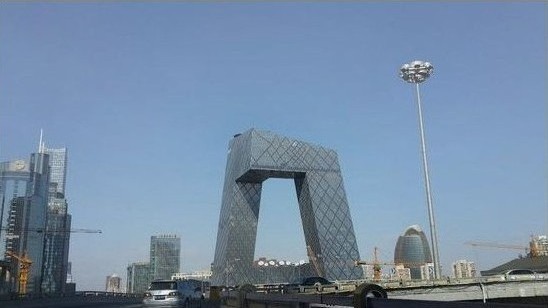}}
  {\includegraphics[width=0.95\textwidth]{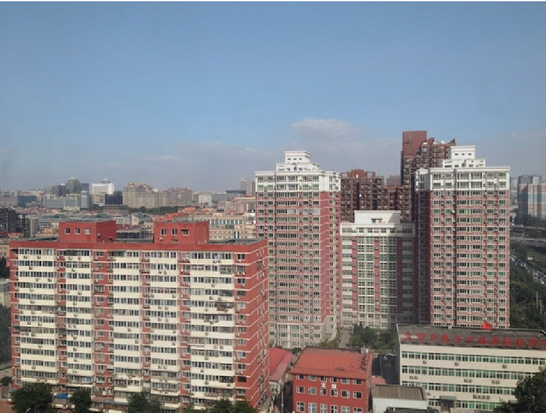}}
    \caption{DehazeNet}
    \label{fig:mscnn}
  \end{subfigure}
\hspace*{-0.04in}%
  \begin{subfigure}{0.142\textwidth}
  \centering
          \fourobjects
  {\includegraphics[width=0.95\textwidth]{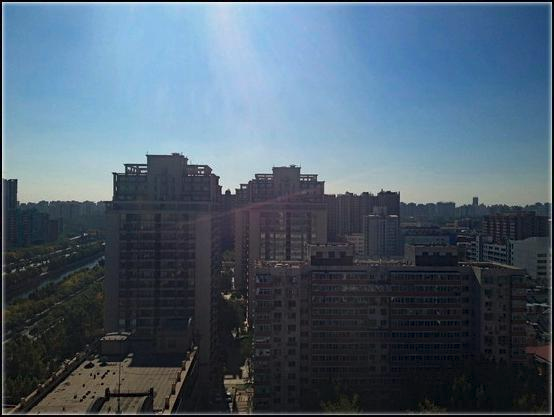}}
  {\includegraphics[width=0.95\textwidth]{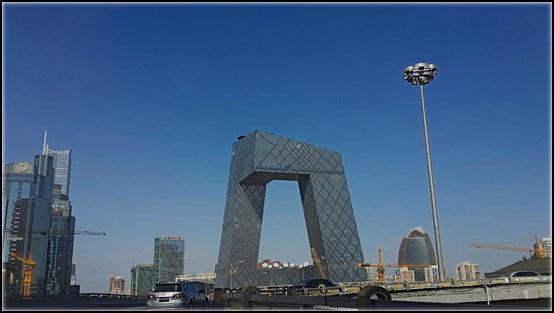}}
  {\includegraphics[width=0.95\textwidth]{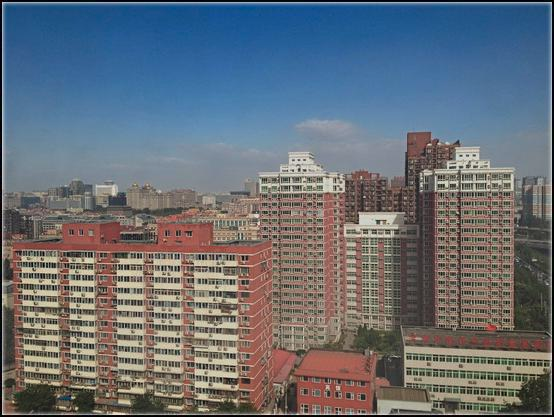}}
    \caption{AOD-Net}
    \label{fig:aodnet}
  \end{subfigure}
\hspace*{-0.04in}%
  \begin{subfigure}{0.142\textwidth}
  \centering
          \fourobjects
  {\includegraphics[width=0.95\textwidth]{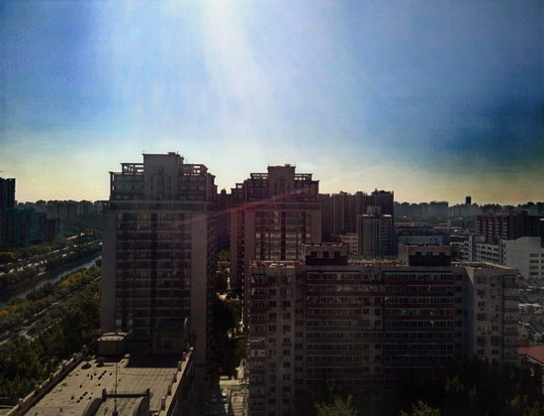}}
  {\includegraphics[width=0.95\textwidth]{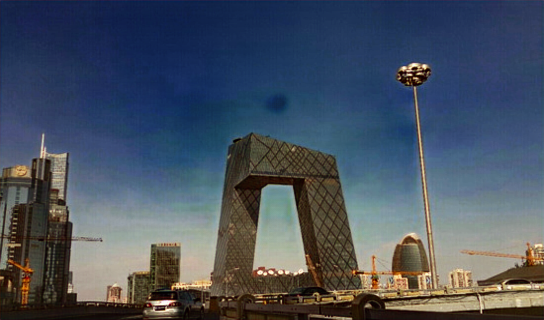}}
  {\includegraphics[width=0.95\textwidth]{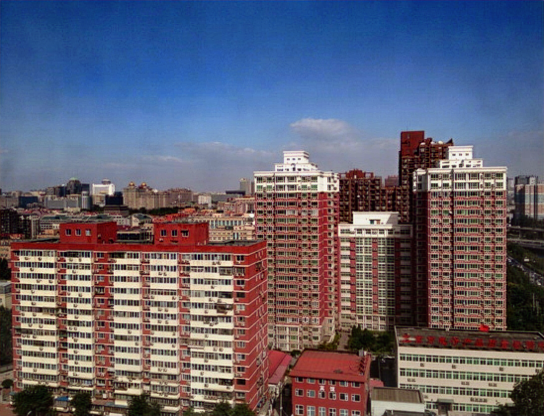}}
    \caption{EPDN}
    \label{fig:dcpdn}
  \end{subfigure}
 \hspace*{-0.04in}%
  \begin{subfigure}{0.142\textwidth}
  \centering
          \fourobjects
  {\includegraphics[width=0.95\textwidth]{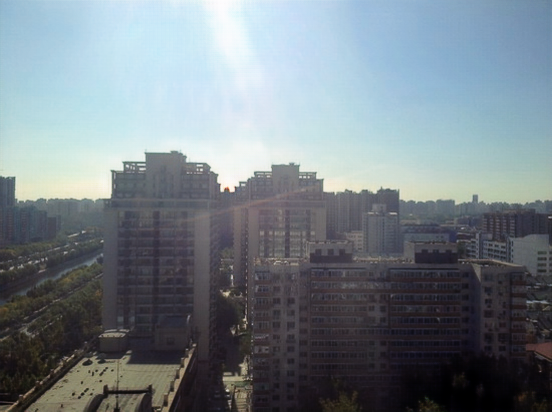}}
  {\includegraphics[width=0.95\textwidth]{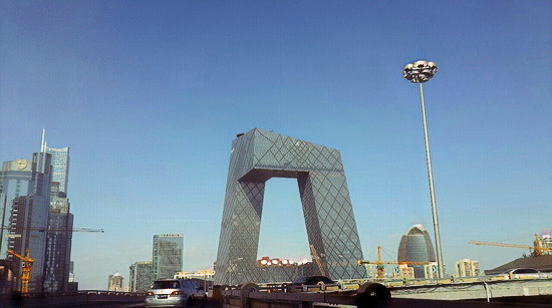}}
  {\includegraphics[width=0.95\textwidth]{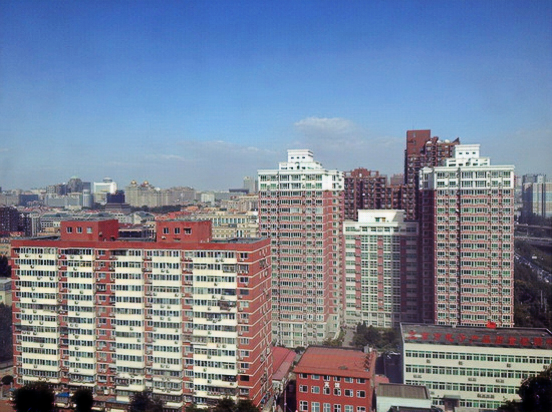}}
    \caption{Ours}
    \label{fig:ours}
  \end{subfigure}
  \hspace*{-0.04in}%
  \begin{subfigure}{0.142\textwidth}
  \centering
          \fourobjects
  {\includegraphics[width=0.95\textwidth]{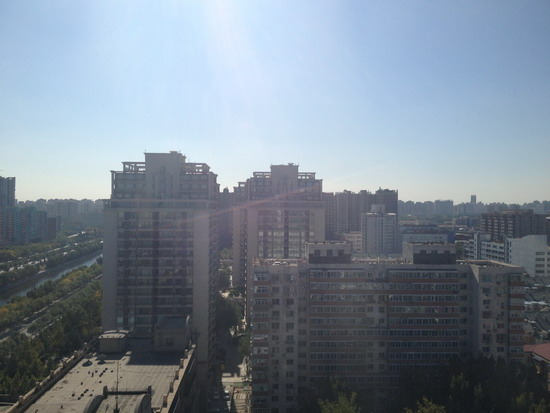}}
  {\includegraphics[width=0.95\textwidth]{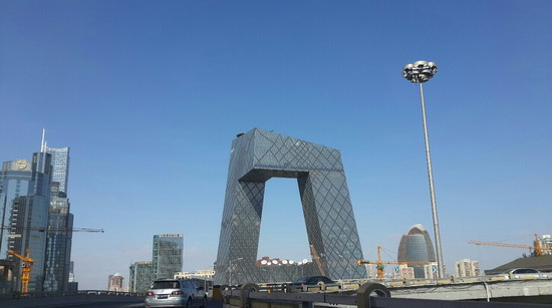}}
  {\includegraphics[width=0.95\textwidth]{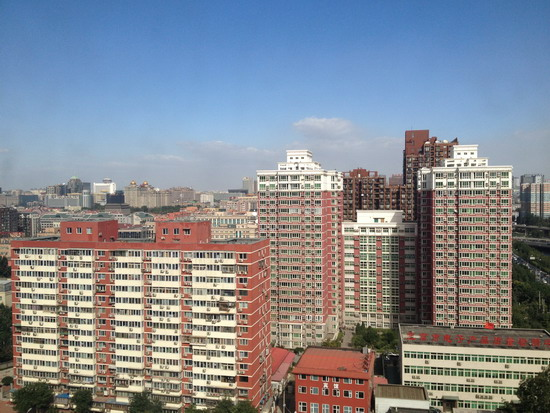}}
    \caption{Haze free}
    \label{fig:gt}
  \end{subfigure}
  \caption{Comparison of the state-of-the-art dehazing methods on SOTS dataset. For high resolution version of these images, please refer to the supplementary material.}
  \label{fig:examples_sots}
\end{figure*}

Figure~\ref{fig:examples_nyu} depicts the qualitative results of our proposed method vs. other single image dehazing methods on NYU-Depth images. As one can see, our method removes more haze as shown in the red boxes. In addition, Figure~\ref{fig:examples_sots} depicts the qualitative results of our proposed method vs. other single image dehazing methods on SOTS images. Our method generates more visually pleasing images while effectively removing haze compared to to other methods.

DCP which is one of the earliest prior-based methods suffers from color distortion and over-exposure. Cycle-Dehaze and CycleGAN fail to remove much haze from dense hazy images. AOD-Net and DehazeNet fail to remove much haze from dense hazy images. EPDN adds generates much darker images compared to the ground truth. DCP suffers from color distortion and over-exposure. DCPDN also suffers from over-exposure. 

Our method, Dehaze-GLCGAN, on the other hand generates more natural and visually pleasant haze-free images and much closer to the ground truth image. Moreover, our method outperforms the above-mentioned methods in recovery of details, and generates more natural images.

\section{Conclusion}
In this paper, we treated the image dehazing problem as an image-to-image translation problem, and proposed a cycle-consistent generative adversarial network, called Dehaze-GLCGAN, for unpaired image dehazing. Dehaze-GLCGAN utilizes discriminators with a local-global structure to remove haze effectively and perceptual and color loss to generate realistic clean images. Using three benchmark test datasets, we showed the effectiveness of the proposed method.

We showed that the  global-local discriminators structure as well as perceptual and color loss can be effectively applied to unpaired single image dehazing through adversarial training. We speculate that this structure can be generalized to other image restoration and reconstruction applications such as single image de-raining.


\Urlmuskip=0mu plus 1mu\relax
\bibliographystyle{acm}
\bibliography{glcgan2020}

\end{document}